\documentclass[fleqn,usenatbib]{mnras}

\usepackage{newtxtext,newtxmath}

\usepackage[T1]{fontenc}
\usepackage{orcidlink}

\DeclareRobustCommand{\VAN}[3]{#2}
\let\VANthebibliography\thebibliography
\def\thebibliography{\DeclareRobustCommand{\VAN}[3]{##3}\VANthebibliography}

\usepackage{graphicx}	
\usepackage{amsmath}	
\usepackage{bm}	
\usepackage{amsbsy}	
\usepackage{xcolor}
\usepackage{multirow}

\newcommand{\lya}{Ly~$\alpha$}
\newcommand{\lyb}{Ly~$\beta$}
\newcommand{\msolh}{\ensuremath{h^{-1}\,{\rm M}_\odot}}
\newcommand{\mpch}{\ensuremath{h^{-1}\,{\rm Mpc}}}
\newcommand{\cmpch}{\ensuremath{h^{-1}\,{\rm cMpc}}}
\newcommand{\xhi}{\ensuremath{x_{\rm HI}}}

\newcommand{\ud}{{\rm d}}
\newcommand{\angstrom}{\text{\normalfont\AA}}

\newcommand{\ieug}{ion.\ eq., uniform $\tilde{\Gamma}_{\rm HI}$}
\newcommand{\ieugtd}{ion.\ eq., uniform $\tilde{\Gamma}_{\rm HI}$, $T(\Delta)$}
\newcommand{\rdfm}{\ensuremath{r(\delta_{F,{\rm max}})}}
\newcommand{\hi}{H\,\textsc{i}}
\newcommand{\civ}{C\,\textsc{iv}}
\newcommand{\oiii}{[O\,\textsc{iii}]}
\newcommand{\eref}[1]{Eq.~\eqref{#1}}
\newcommand{\sref}[1]{Sec.~\ref{#1}}
\newcommand{\fref}[1]{Fig.~\ref{#1}}
\newcommand{\tref}[1]{Table~\ref{#1}}

\title[The galaxy--\lya{} transmission connection]{The connection between high-redshift galaxies and Lyman {\boldmath $\alpha$} transmission in the Sherwood-Relics simulations of patchy reionisation}

\author[L. Conaboy et al.]{Luke Conaboy$^{1}$\,\orcidlink{0000-0002-6580-7177}\thanks{E-mail: luke.conaboy@nottingham.ac.uk},
  James S. Bolton$^{1}$\,\orcidlink{0000-0003-2764-8248},
  Laura C. Keating$^{2}$\,\orcidlink{0000-0001-5211-1958}, 
  Martin G. Haehnelt$^{3}$\,\orcidlink{0000-0001-8443-2393}, 
  Girish Kulkarni$^{4}$\,\orcidlink{0000-0001-5829-4716}
  \newauthor
  and Ewald Puchwein$^{5}$\,\orcidlink{0000-0001-8778-7587}\\
  $^{1}$School of Physics and Astronomy, The University of Nottingham, University Park, Nottingham, NG7 2RD, UK\\
  $^{2}$Institute for Astronomy, University of Edinburgh, Blackford Hill, Edinburgh, EH9 3HJ, UK \\
  $^{3}$Kavli Institute for Cosmology and Institute of Astronomy, Madingley Road, Cambridge, CB3 0HA, UK\\
  $^{4}$Tata Institute of Fundamental Research, Homi Bhabha Road, Mumbai 400005, India\\
  $^{5}$Leibniz-Institut f\"ur Astrophysik Potsdam, An der Sternwarte 16, 14482 Potsdam, Germany\\}

\date{17 April 2025}

\pubyear{2025}

\begin{document}
\label{firstpage}
\pagerange{\pageref{firstpage}--\pageref{lastpage}}
\maketitle

\begin{abstract}
  Recent work has suggested that, during reionisation, spatial
  variations in the ionising radiation field should produce enhanced
  \lya{} forest transmission at distances of tens of comoving Mpc from
  high-redshift galaxies.  We demonstrate that the Sherwood-Relics
  suite of hybrid radiation-hydrodynamical simulations are
  qualitatively consistent with this interpretation. The shape of the
  galaxy--\lya{} transmission cross-correlation is sensitive to both
  the mass of the haloes hosting the galaxies and the volume averaged
  fraction of neutral hydrogen in the IGM, $\bar{x}_{\rm HI}$. The
  reported excess \lya{} forest transmission on scales
  $r \sim 10~{\rm cMpc}$ at $\langle z \rangle\approx 5.2$ -- as measured using \civ{}
  absorbers as proxies for high-redshift galaxies -- is quantitatively
  reproduced by Sherwood-Relics at $z=6$ if we assume the galaxies
  that produce ionising photons are hosted in haloes with mass
  $M_{\rm h}\geq 10^{10}~\msolh$. However, this redshift mismatch is
  equivalent to requiring $\bar{x}_{\rm HI}\sim 0.1$ at $z\simeq 5.2$, which
  is inconsistent with the observed \lya{} forest effective optical
  depth distribution.  We suggest this tension may be partly
  resolved if the minimum \civ{} absorber host halo mass at $z>5$ is
  larger than $M_{\rm h}=10^{10}~\msolh$.  After reionisation
  completes, relic IGM temperature fluctuations will continue to
  influence the shape of the cross-correlation on scales of a few
  comoving Mpc at $4 \leq z \leq 5$.  Constraining the redshift evolution of
  the cross-correlation over this period may therefore provide further
  insight into the timing of reionisation.
\end{abstract}

\begin{keywords}
  methods: numerical -- intergalactic medium -- galaxies: high-redshift
  -- quasars: absorption lines -- large scale structure of Universe --
  dark ages, reionization, first stars
\end{keywords}


\section{Introduction}
\label{sec:int}
Between the first hundred million years and the first billion years,
the hydrogen gas permeating the Universe transitioned from cold and
neutral to warm and (re)ionised. A wide variety of observational constraints place the midpoint of
reionisation between $7 \lesssim z \lesssim 8$ \citep[e.g.,][]{davies2018, banados2018,
  greig2019, jung2020, planckcollaboration2020, gaikwad2023, jin2023,
  durovcikova2024, umeda2024}.   The rapid evolution of the mean free path of Lyman-limit photons
between $5< z <6$ \citep{becker2021, zhu2023, satyavolu2024}, the
distribution of the \lya{} forest transmission \citep{bosman2022},
and the presence of damping wings in the spectra of $z<6$ quasars
\citep{becker2024,spina2024, zhu2024a} are furthermore consistent with the last neutral hydrogen islands persisting to $z < 6$.  This suggests 
reionisation may have finished around $z \approx 5.3$
\citep{kulkarni2019, keating2020, keating2020a, nasir2020}.

Concurrently, {\em JWST} is opening up a new window on galaxy
formation at $z \gtrsim 8$ \citep[e.g.,][]{naidu2022, arrabalharo2023a,
  castellano2024, harikane2024, zavala2024}. Constraints from deep
observations with {\em JWST} suggest that faint, low-mass galaxies are
efficient at producing ionising photons, and can provide the majority
of the ionising photons required for reionisation (\citealt{atek2024,
  saxena2024, simmonds2024, begley2024}, although see also
\citealt{gazagnes2024}).  The discovery of many faint AGN
\citep[e.g.][]{harikane2023} also means that a contribution to
the ionising photon budget by faint AGN is plausible
\citep{maiolino2024, grazian2024, madau2024, asthana2024a}.

A promising route for connecting the properties of the IGM with these
high redshift ionising sources is a cross-correlation with
intergalactic \lya{} transmission. At low redshifts ($z<5$), well
after the end of reionisation, the galaxy--\lya{} transmission
correlation has been extensively studied \citep{adelberger2005,
  tummuangpak2014, turner2014, bielby2017, matthee2024, banerjee2024}.
The low-redshift ($2\lesssim z \lesssim 4$) picture that has emerged is one of a
strong decrease in \lya{} transmission approaching galaxies, driven by
enhanced overdensities and the clustering and infall of neutral
hydrogen in the regions where these galaxies reside. Models of this
low-redshift deficit in transmission at small impact parameters
($b\lesssim 1~{\rm cMpc}$) also suggest that stellar feedback plays an
important role in setting the level of \lya{} transmission
\citep{meiksin2017, sorini2020}.

At higher redshifts, the picture is less clear. \citet{kakiichi2018}
used a small sample of spectroscopically-confirmed Lyman break
galaxies (LBGs) at $5.3 < z < 6.4$ in the field of a $z=6.42$ QSO to
examine the impact of local ionisation on \lya{} transmission. They
cross-correlated the positions of these LBGs with the \lya{} forest,
finding that there was some evidence (tempered by the small sample
size) for an enhancement of \lya{} transmission in proximity to
galaxies. This enhancement they attributed not solely to the observed
LBGs, but also to the ionising radiation produced by nearby, faint,
and therefore undetected, sources. \citet{meyer2019} computed the 1D
correlation between \civ{} absorbers and \lya{} transmission in the
spectra of 25 QSOs at $4.5 < z < 6.3$, finding reduced transmission at
$r\lesssim 7~{\rm cMpc}$ (qualitatively similar to the low-redshift results
discussed previously, though quantitatively the deficit in
transmission in \citealt{meyer2019} is somewhat stronger) and
statistically-significant enhanced \lya{} transmission relative to the
mean at distances of $r\sim 15$--$45~{\rm cMpc}$. In a follow-up work,
\citet{meyer2020} characterised the galaxy--\lya{} transmission
correlation at $z\sim 6$, but were unable to reproduce the findings of
\citet{meyer2019}, which they attribute to a small sample size and
noise. Additionally, \citet{kashino2023} reported a measurement of the galaxy--\lya{} transmission correlation from a single QSO sightline
from the EIGER survey at $5.3< z < 5.7$ and $5.7< z <
6.14$. Finally, while this work was being completed, measurements of the galaxy--\lya{} transmission correlation over the redshift range $5.4<z<6.5$ from the {\em JWST} ASPIRE survey were released \citep{kakiichi2025}.  Using five QSO fields, these authors reported $2\sigma$ evidence for excess \lya\ transmission on scales of $20$--$40\rm\,cMpc$ from \oiii\ emitting galaxies at $\langle z \rangle = 5.86$.

Modelling of the high-redshift galaxy--\lya{} transmission correlation
using reionisation simulations has also returned a variety of
results. \citet{garaldi2019} performed the first examination of the
transmission correlation in the high-redshift context, using the CROC
simulations of cosmic reionisation
\citep{gnedin2014}. \citet{garaldi2019} compared the results of the
CROC simulation to the \citet{meyer2019} observational constraints,
finding that their models show a reduction in transmission close to
galaxies (in a similar fashion to the low-$z$ observations) but do not
reproduce the enhanced transmission at larger radii reported by
\citet{meyer2019}. \citet{zhu2024} performed a further analysis of the
CROC simulations and confirmed the absence of an excess in
transmission. In contrast, \citet{garaldi2022} showed that, in the
THESAN simulations \citep{kannan2022, smith2022}, the galaxy--\lya{}
transmission correlation shows both reduced transmission close to
galaxies and excess transmission at larger distances.  This is
qualitatively similar to \citet{meyer2019}, although good agreement
with the \citet{meyer2019} results occurs at a different (higher)
redshift in the models.  \citet{garaldi2022} also demonstrated that,
in the context of the THESAN model, the galaxy--\lya{} transmission
correlation is largely independent of the source model when
reionisation history is accounted for (e.g., by comparing the
correlation at fixed neutral fraction $x_{\rm HI}$, as opposed to
fixed redshift). Recently, \citet{garaldi2024} performed a deeper
analysis of the galaxy--\lya{} transmission correlation in THESAN,
focusing on its observability.  They noted that the selection criteria
for galaxies (e.g., \civ{} absorption, \oiii{} flux) appears to have
little impact on the correlation in their models.

In this context, we examine the galaxy--\lya{} transmission  correlation in the
Sherwood-Relics\footnote{\url{https://www.nottingham.ac.uk/astronomy/sherwood-relics/}}
simulation suite \citep{puchwein2023}.  In this
work we focus on the high-redshift ($z\geq 4.2$) galaxy--\lya{}
transmission correlation, at distances $r\gtrsim {\rm few~cMpc}$ from
galaxies. Sherwood-Relics has
demonstrated excellent agreement with observations for a range of IGM
properties at $z\lesssim 6$ \citep[see e.g.,][]{gaikwad2020, molaro2022,
  feron2024}. The simulations differ from other numerical models in several ways.  First, they have been designed specifically for modelling the properties of the \lya{} forest at the tail end of reionisation,
making them ideal for investigating the galaxy--\lya{} transmission
correlation at $z\sim 6$. Second, instead of full radiation-hydrodynamics, Sherwood-Relics uses a novel
hybrid radiative transfer (RT) scheme (see \sref{sec:mod-she}) that
captures the hydrodynamical response of the gas to the inhomogeneous
photoheating and photoionisation associated with reionisation.  This allows different reionisation histories to be examined at relatively low computational cost.

The paper is structured as follows.  In \sref{sec:mod-she} we briefly describe
the simulations used in this work, and in \sref{sec:mod-rea} we perform an initial examination of the \lya{} transmission around galaxies in our fiducial model to gain intuition.  Next, in
\sref{sec:res} we explore the galaxy--\lya{} transmission correlation
in Sherwood-Relics, and in \sref{sec:res-phy-main} we explore the effect of different modelling assumptions in an attempt to elucidate the physical origin
of the galaxy--\lya{} transmission correlation, both during and after reionisation.  Finally, we conclude in \sref{sec:con}.  We assume a
flat $\Lambda$CDM cosmology with $\Omega_\Lambda=0.692$,
$\Omega_{\rm m}=0.308$, $\Omega_{\rm b}=0.0482$, $\sigma_8=0.829$,
$n_s=0.961$ and $h=0.678$ \citep{planckcollaboration2014}. Unless
otherwise specified, distance units are comoving (and may be
explicitly specified as such by the prefix `c').


\section{Modelling {\boldmath\lya{}} transmission}
\label{sec:mod}

\subsection{Sherwood-Relics simulations}
\label{sec:mod-she}
The simulations used in this work form part of the Sherwood-Relics
project \citep[see][for details]{puchwein2023}. The Sherwood-Relics
suite of simulations are a series of high-resolution cosmological
hydrodynamical simulations performed using a modified version of {\sc
  p-gadget-3} (itself a modified version of {\sc gadget-2}, described
in \citealt{springel2005}). In this study, we use runs with box sizes
of $40~\mpch$ (40-2048) and $160~\mpch$ (160-2048), each containing
$2 \times 2048^3$ particles. In addition, for testing the convergence of
our results with mass resolution, we also employ variants of the
$40~\mpch$ box that are identical to 40-2048, except they were run
with $2\times1024^3$ (40-1024) and $2\times 512^3$ particles (40-512). See
\tref{tab:mod-run} for a summary of the runs used in this work.
\begin{table*}
  \caption{Overview of the Sherwood-Relics runs employed in this study
    \citep[see also][]{puchwein2023}. Listed in columns are: the name
    used to refer to the run; linear comoving box size in $\cmpch$
    ($L_{\rm box}$); total initial number of dark matter and gas
    particles ($N_{\rm part}$); dark matter particle mass in $\msolh$
    ($M_{\rm d}$); gas particle mass in $\msolh$ ($M_{\rm g}$);
    redshift at which the global volume-averaged neutral fraction
    drops below $10^{-3}$ ($z_{\rm r}$), and the output
    redshifts for snapshots $z_{\rm snap}$. The lowest redshift
    snapshot for all models is $z=4.2$.}
\label{tab:mod-run}
\begin{tabular}{lcccccccc}
\hline
Name & $L_{\rm box}$ ($\cmpch$) & $N_{\rm part}$ & $M_{\rm d}$ $(\msolh)$ & $M_{\rm g}$ $(\msolh)$ & $z_{\rm r}$ & $z_{\rm mid}$ & $z_{\rm snap}$\\
\hline
  40-2048 (fiducial) & 40  & $2 \times 2048^3$ & $5.37\times10^5$ & $9.97\times10^4$ & $5.7$ & 7.5 & every $\Delta z =0.2$\\
  40-2048 (late)     & 40  & $2 \times 2048^3$ & $5.37\times10^5$ & $9.97\times10^4$ & $5.3$ & 7.2 & $z=7, 6, 5.4, 4.8,4.2$\\  
  40-2048 (mid)      & 40  & $2 \times 2048^3$ & $5.37\times10^5$ & $9.97\times10^4$ & $6.0$ & 7.3 & $z=7, 6, 5.4, 4.8,4.2$\\  
  40-2048 (early)    & 40  & $2 \times 2048^3$ & $5.37\times10^5$ & $9.97\times10^4$ & $6.6$ & 8.0 & $z=7, 6, 5.4, 4.8,4.2$\\  
  40-1024            & 40  & $2 \times 1024^3$ & $4.30\times10^6$ & $7.97\times10^5$ & $5.7$ & 7.5 & every $\Delta z=0.2$ \\
  40-512             & 40  & $2 \times 512^3$  & $3.44\times10^7$ & $6.38\times10^6$ & $5.7$ & 7.5 & every $\Delta z=0.2$ \\
  160-2048           & 160 & $2 \times 2048^3$ & $3.44\times10^7$ & $6.38\times10^6$ & $5.3$ & 7.2 & every $\Delta z=0.2$ \\
\hline
\end{tabular}
\end{table*}

The Sherwood-Relics project uses a novel hybrid radiative transfer
(RT) scheme to capture the hydrodynamical effects of inhomogeneous
reionisation, without incurring the expense of fully-coupled
radiation-hydrodynamical (RHD) simulations. Here we briefly outline
key points of the method, but for the full exposition see
\citet{puchwein2023}. In this scheme, the monochromatic radiative
transfer of UV photons is followed using the moment-based, M1-closure
RT code {\sc aton} \citep{aubert2008}. {\sc aton} is run in
post-processing mode (i.e. on a periodically refreshed, rather than
continuously evolving, density field, meaning that the evolution of
the density is not coupled to that of the radiation) on a base {\sc
  p-gadget-3} simulation, with input snapshots spaced every
$\Delta t=40~{\rm Myr}$. In order to use the density fields from
  the base {\sc p-gadget-3} run with {\sc aton}, the gas particles are first deposited on a uniform grid using the smoothed particle hydrodynamics (SPH) kernel. From this
{\sc aton} simulation, we extract three-dimensional maps of the
inhomogeneous \hi{} photoionisation rate $\Gamma_{\rm HI}$ at each
redshift. These $\Gamma_{\rm HI}$ maps then serve as inputs to a new {\sc
  p-gadget-3} simulation, thus allowing the hydrodynamic response of
the gas to the spatially-varying radiation field to be captured
without running expensive RHD simulations. The luminosity of an
ionising source is assumed to be proportional to its halo mass, and
the minimum mass of ionising sources is $M_{\rm h} >
10^9~\msolh$. Ionising photons have mean energy $18.6~{\rm eV}$, which
corresponds to a blackbody spectrum with temperature
$T=4\times10^4~{\rm K}$. The exact emissivity of each halo is not a
prediction but is determined by fixing the redshift evolution of the
total ionising emissivity, thus allowing the exact reionisation
history to be calibrated \citep{kulkarni2019, keating2020}.

\begin{figure*}
  \centering
  \includegraphics[width=\linewidth]{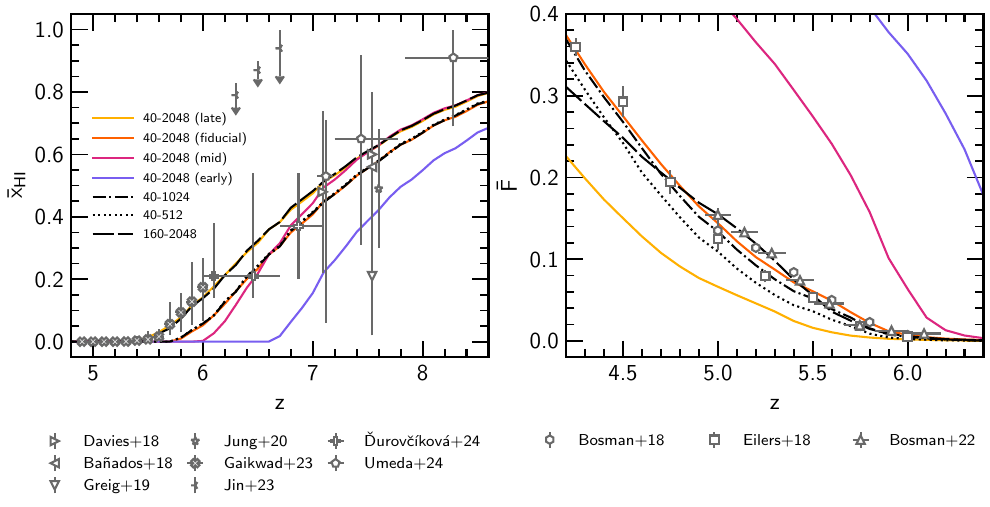}
  \vspace{-0.5cm}
  \caption{Left: evolution of the volume-averaged neutral fraction
    $\bar{x}_{\rm HI}$ in the 40-2048 (coloured solid), 40-1024
    (dot-dashed), 40-512 (dotted) and 160-2048 (long-dashed) runs. For
    the 40-2048 run we show four different reionisation models: late
    (gold), fiducial (orange), mid (dark pink) and early (purple) --
    see \tref{tab:mod-run} for details. Also shown are observational
    constraints on $\bar{x}_{\rm HI}$ derived from: the effective
    optical depth of the \lya{} forest \citep{gaikwad2023}; quasar
    damping wings \citep[][shown for the {\sc aton} model
    constraints]{banados2018, davies2018, greig2019, durovcikova2024};
    the \lya{} emitter equivalent width distribution \citep{jung2020};
    \lya{} and \lyb{} dark pixel fractions \citep{jin2023}; and galaxy
    damping wings \citep{umeda2024}. Right: the mean transmission in
    the \lya{} forest over 50~\cmpch{} skewers, with constraints from
    \citet{bosman2018, eilers2018, bosman2022}. The fiducial 40-2048
    model is calibrated to reproduce the \citet{bosman2018} and
    \citet{eilers2018} results, while the 160-2048 model is calibrated
    to reproduce the \citet{bosman2022} results.}
  \label{fig:xHI-all}
\end{figure*}

In \fref{fig:xHI-all} we show the redshift evolution of the
volume-averaged neutral fraction $\bar{x}_{\rm HI}$ and the mean
transmission, $\bar{F}$, of the \lya{} forest in 50~\cmpch{} skewers
for all the models used in this work. Note that the 40-1024 and 40-512
models use the same reionisation history as the fiducial 40-2048
model. Two of the models we use are calibrated to match measurements
of $\bar{F}$: the fiducial 40-2048 model is calibrated to the
\citet{bosman2018} and \citet{eilers2018} measurements, while the
160-2048 model is calibrated to match the \citet{bosman2022}
measurements. In addition to the calibrated reionisation histories, we
also use three other models where reionisation ends at:
$z_{\rm r}=5.3$ (late), $z_{\rm r}= 6.0$ (mid), and $z_{\rm r}=6.7$
(early). These models are not calibrated to measurements of $\bar{F}$,
but are useful for observing the effect of an altered reionisation
history.  Note also that despite using the same reionisation history
as the fiducial 40-2048 model, differences in mass resolution mean
that the 40-1024 and 40-512 models do not exactly match the
\citet{bosman2018} and \citet{eilers2018} results.

The Sherwood-Relics project is primarily concerned with modelling the
high-redshift IGM, and so does not include detailed subgrid physics
models for galaxy formation. Instead, a computationally-efficient
density removal scheme -- the `quick \lya{}' approach -- is employed,
where gas particles with overdensity
$\rho_{\rm g}/\bar{\rho}_{\rm g} = \Delta > 1000$ and temperature
$T<10^5~{\rm K}$ are converted to collisionless star particles
\citep{viel2004}. The simulations do not include any prescription for
stellar feedback. Halo finding is performed using the inbuilt
friends-of-friends (FoF) halo finder with a linking length of 0.2
times the mean interparticle spacing \citep{springel2005}. When we
discuss `halo masses', we are referring to the total FoF group mass
$M_{\rm h} = M_{\rm h,d}+ M_{\rm h,g} + M_{\rm h, \star}$, where
$M_{\rm h,d}$, $M_{\rm h,g}$ and $M_{\rm h, \star}$ are the dark matter,
gas and stellar components of the FoF group, respectively.

Sightlines are drawn through the simulation volume to extract the
relevant quantities for computing mock \lya{} absorption spectra. We
draw 5000 such sightlines, with a spatial resolution of
$28.8~{\rm ckpc}$ ($115.2~{\rm ckpc}$) for the $40~\mpch$
($160~\mpch$) box. These sightlines are then post-processed to compute
the \lya{} optical depths, using the approximation to the Voigt line
profile due to \citet{tepper-garcia2006} and including, unless
otherwise stated, the effects of peculiar velocities. Note
  that we do not attempt to model observational effects (e.g., noise,
  spectral resolution, total spectrum length) in these mock spectra,
  instead deferring a study of these effects for future work
  \citep[see also][for a recent study exploring the impact of these
  effects]{garaldi2024}.

\subsection{Real-space {\boldmath \lya{}} transmission}
\label{sec:mod-rea}
Before proceeding further, to develop some intuition about the nature of the galaxy--\lya{} transmission
correlation, following \citet{kulkarni2015} we consider the real-space \lya{} transmission.  We will return to examining the full 
line-of-sight calculation of \lya{} optical depths, including redshift space distortions, in \sref{sec:res}.  The real-space \lya{} transmission is defined as
\begin{equation}
  \label{eq:sim-Freal}
  F_{\rm real} = \operatorname{exp}\left[- \frac{3 \lambda_\alpha^3 \Lambda_\alpha}{8\pi H(z)}x_{\rm HI}\Delta{\bar n}_{\rm H} \right],
\end{equation}
where $\lambda_\alpha=1216~\angstrom$ is the rest-frame \lya{} wavelength,
$\Lambda_\alpha=6.265\times10^{-8}{~\rm s^{-1}}$, $H(z)$ is the Hubble parameter, ${\bar n}_{\rm H}$ is the mean number density of hydrogen, $\Delta$ is the overdensity and $x_{\rm HI}$ is the neutral hydrogen fraction, which depends on both the photoionisation rate, $\Gamma_{\rm HI}$, and the gas temperature, $T$.  Note that
\eref{eq:sim-Freal} is equal to $\operatorname{exp}(-\tau_{\rm GP})$,
where $\tau_{\rm GP}$ is the Gunn--Peterson optical depth \citep{gunn1965}
and, unlike the calculations performed in the rest of this work,
\eref{eq:sim-Freal} ignores the effect of peculiar velocities and
thermal line broadening on the \lya{} optical depth, and assumes that
the line profile is a Dirac delta function at
$\lambda=\lambda_{\alpha}$.

\begin{figure*}
  \centering
  \includegraphics[width=\textwidth]{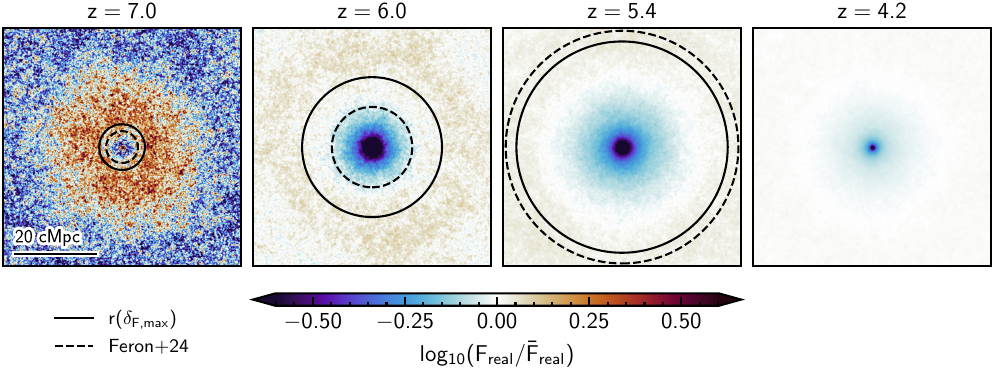}
  \vspace{-0.5cm}
  \caption{Logarithm of the normalised real-space \lya{} transmission
    in stacked slices, each centred on a halo with mass
    $M_{\rm h} \geq 10^{10}~\msolh$ for the fiducial 40-2048 model. Each
    slice has width $40~\cmpch$ and depth $115.2~{\rm ckpc}$. We show
    the stacked transmission fluctuation at fixed redshifts of (from
    left to right) $z=7.0$, $6.0$, $5.4$ and $4.2$, where each stack
    is produced from 2,431, 4,540, 6,191 and 10,706 slices,
    respectively. To highlight the structure of the transmission
    correlation, we clip the dynamic range of the colour scale
    to be from $\bar{F}_{\rm real}/4$ to
      $4 \bar{F}_{\rm real}$, as indicated by the arrows on the
    colour bar. Also shown for comparison (see \sref{sec:res-gal} for
    a discussion) is the distance at which the halo--\lya{}
    transmission correlation (calculated without including peculiar
    velocities) is largest (`$r(\delta_{F,{\rm max}})$', black solid
    circle), and the mean free path of Lyman limit photons around
    haloes from \citet{feron2024} (`$\Lambda_{\rm mfp}$', black dashed
    circle).}
  \label{fig:sim-proj}
\end{figure*}

\begin{figure*}
  \centering
  \includegraphics[width=\linewidth]{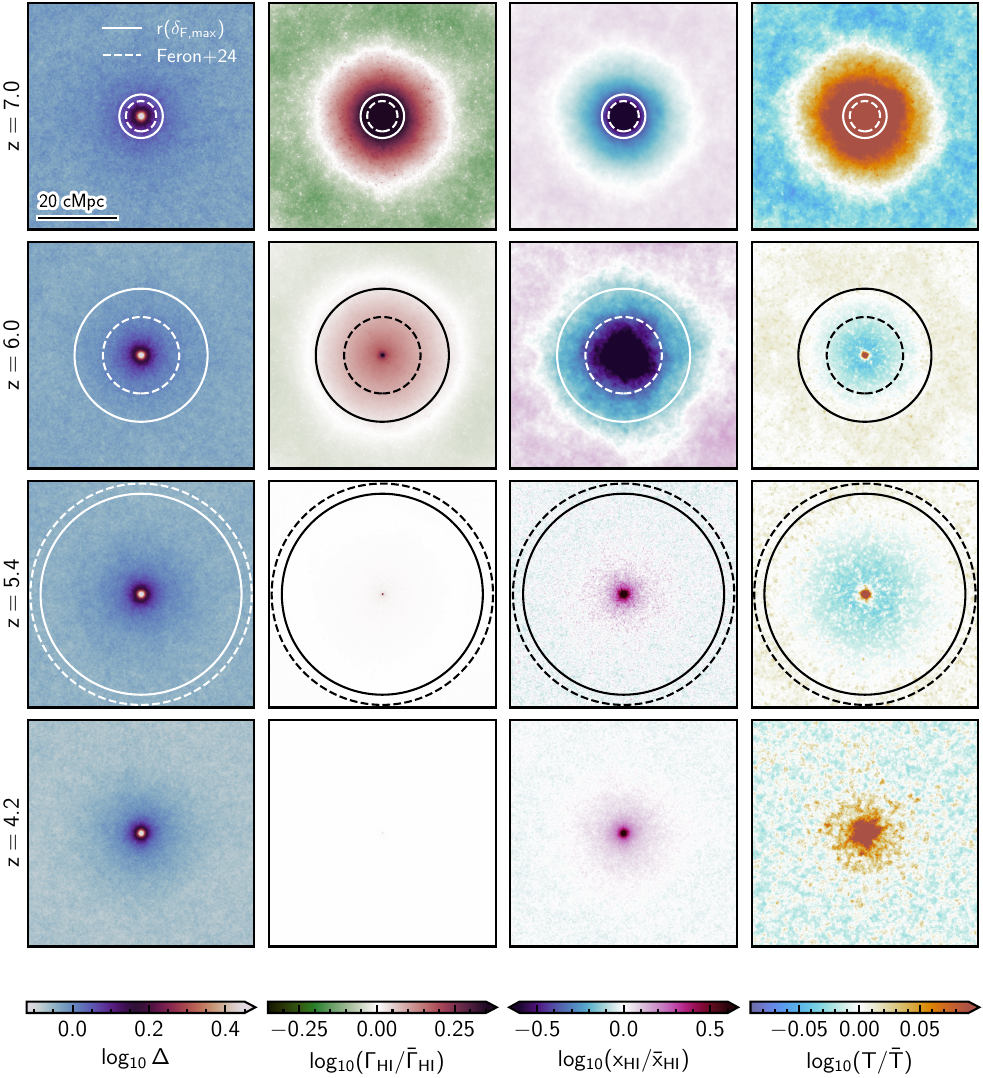}
  \vspace{-0.5cm}
  \caption{Stacked slices of the key quantities that set the
    real-space \lya{} transmission displayed in \fref{fig:sim-proj}
    (the slices used to produce the stacks are the same as those used
    to make \fref{fig:sim-proj}). From left to right, the columns show
    the fluctuations in: gas density
    $\log_{10}\Delta=\log_{10}(\rho_{\rm g}/\bar{\rho}_{\rm g})$; hydrogen
    photoionisation rate
    $\log_{10}(\Gamma_{\rm HI}/\bar{\Gamma}_{\rm HI})$; neutral hydrogen
    fraction $\log_{10}(x_{\rm HI}/\bar{x}_{\rm HI})$; and gas
    temperature $\log_{10}(T/\bar{T})$. In each case, barred
    quantities indicate the global volume average of a quantity. From
    top to bottom, the rows display the stacks for $z=7,6,5.4$ and
    $4.2$. Note that in each case we clip the dynamic range of the
    colourmap, in order to centre the mean and highlight structure
    (the clipped end of the colourbar is triangular). As in
    \fref{fig:sim-proj}, we also show the distance at which the
    halo--\lya{} transmission correlation is largest
    (`$r(\delta_{F,{\rm max}})$', solid circle), and the mean free path of
    Lyman limit photons around haloes from \citet{feron2024} (dashed
    circle).}
  \label{fig:sim-proj-qty}
\end{figure*}

We calculate $F_{\rm real}$ by first depositing the particles in the
simulation onto a uniform grid using the SPH kernel; the \lya{} transmission can then be computed for each pixel independently.  Next, we must select the haloes that host galaxies in the simulation.  Measurements of the galaxy--\lya{} transmission correlation often use different methods for the detection of galaxies, such as \civ{} absorption or \oiii{}
emission. Given these differing detection methods, it is not
immediately clear how to relate the sources of absorption in the
observations to the structures in our simulations. \citet{meyer2019}
used abundance matching arguments to relate the incidence of \civ{}
absorbers per absorption distance $\ud \mathcal{N}/\ud X$ to the halo mass
function and thus obtain an approximate lower mass limit on the haloes
hosting these absorbers, finding $M_{\rm h}\gtrsim 10^{9.8}~\msolh$.  Their
measurement of the \civ{} absorber bias is very uncertain, however,
and is consistent with host halo masses as large as
$M_{\rm h}\sim 10^{12.4}~\msolh$.  \citet{pizzati2024} use the FLAMINGO
simulations \citep{schaye2023} to reproduce observations of the
cross-correlation of high-$z$ galaxies (selected through \oiii{}
emission), finding that the characteristic host halo mass of an
\oiii{} emitter is $\approx 10^{10.7}~\msolh$. However, \citet{garaldi2022}
and \citet{garaldi2024} found that the exact choice of selection
criteria (e.g. \civ{} column density or halo mass) makes only a small
impact on the galaxy--\lya{} transmission correlation. Motivated by
these arguments, here and throughout the rest of this work we
therefore compute the galaxy--\lya{} transmission correlation for
haloes with $M_{\rm h} \geq 10^{10}~\msolh$, unless otherwise specified.
Finally, we extract a slice of depth $115.2~{\rm ckpc}$ for each halo
with $M_{\rm h}\geq 10^{10}~\msolh$ in the fiducial 40-2048 simulation,
by recentering the grids on the centre of mass of every halo. We stack
slices to compute the average value of $F_{\rm real}$ in each pixel,
before normalising by the global volume-averaged transmission
$\bar{F}_{\rm real}$.

\fref{fig:sim-proj} displays
the normalised stacked transmission $F_{\rm real}/\bar{F}_{\rm real}$
at $z=7,6,5.4~{\rm and}~4.2$. We show the result 
for the fiducial 40-2048 simulation, but note that the general features are the same for the other reionisation models (aside from
differences in timing due to the different reionisation
histories).  At all redshifts shown in \fref{fig:sim-proj}, there is a central
region ($\sim$~few~cMpc in extent) where
$\log_{10}(F_{\rm real}/{\bar F}_{\rm real}) \ll 0$ (i.e. the \lya{} transmission is
lower than average).
At $z=7$, this region of low transmission is followed by an annulus of
excess transmission (i.e.
$\log_{10}(F_{\rm real}/{\bar F}_{\rm real}) > 0$), extending out to a
radius of $\sim 10{~\rm cMpc}$. Beyond this region, the transmission
again drops below the average. The picture is similar for $z=6$ and
$z=5.4$, but the central reduction in transmission is stronger, and
the excess transmission begins at a larger distance from the halo, extends further, and is
weaker. By $z=4.2$ the central region of low transmission is the only
feature that is still visible in the map.

The behaviour of the transmission 
in \fref{fig:sim-proj} may be understood by examining \fref{fig:sim-proj-qty}, where we show averages of
the key quantities used in \eref{eq:sim-Freal}, namely gas overdensity $\Delta$,
photoionisation rate $\Gamma_{\rm HI}$, neutral hydrogen fraction $x_{\rm HI}$ and gas
temperature $T$.\footnote{Note that \fref{fig:sim-proj-qty} shows averages
  over each quantity relevant to \eref{eq:sim-Freal}, \emph{not} the
  values of each quantity that will produce the average $F_{\rm real}$
  shown in \fref{fig:sim-proj}.}  The stacks in \fref{fig:sim-proj-qty} are calculated in the same
fashion as in \fref{fig:sim-proj}.  At all redshifts, $\Delta$ is centrally concentrated with a value at
$r\gtrsim20~{\rm cMpc}$ that gradually becomes smaller with time as structure formation progresses.  At $z=7$, the local radiation field around the haloes is clearly visible in the excess $\Gamma_{\rm HI}$ at $r\lesssim 15~{\rm cMpc}$.  This drives a corresponding decrease in $x_{\rm HI}$ and
excess in $T$ due to photoionisation and photoheating, respectively (i.e., from the ionised bubbles around the sources).  This is also demonstrated by the dashed circles in  and \fref{fig:sim-proj} and  \fref{fig:sim-proj-qty} which show the Lyman-limit mean free path around haloes, $\Lambda_{\rm mfp}$, from \citet{feron2024}.\footnote{The
halo mass bins used to calculate $\Lambda_{\rm mfp}$ in \citet{feron2024} are slightly different to
those employed here, with a mean halo mass
$\langle M_{\rm h}/\msolh\rangle$ of $10^{10.5}$ at $z=7$, $10^{10.7}$ at
$z=6$ and $10^{10.9}$ at $z=5.4$.}  This implies the local radiation field from ionising sources plays an important role in setting the galaxy--\lya{} transmission correlation.  We return to discuss the solid circles in \fref{fig:sim-proj} and \fref{fig:sim-proj-qty} in \sref{sec:res-gal}.

Toward lower redshift the fluctuations in
$\Gamma_{\rm HI}$ fade away as the intensity of the UV background (UVB) increases, and by $z=5.4$ only the very central
excess is visible. The effect of the diminishing UVB fluctuations are again mirrored in the behaviour of the neutral hydrogen fraction (and hence also $F_{\rm real})$.  Lastly, at $z=4.2$ (after the end of reionisation) all the gas in the IGM is highly ionised and the central
region in the stacked slice is more neutral than average due to the larger gas density (i.e., $x_{\rm HI} \propto \Delta$ for gas in ionisation equilibrium).

The behaviour of the temperature fluctuations, displayed in the right
column of \fref{fig:sim-proj-qty}, is slightly different. The central
region at all redshifts contains hot gas associated with shocks from
structure formation. At $z=7$, the region around the stacked haloes
($r\lesssim 15~{\rm cMpc}$) is dominated by recently photoheated gas, which
produces an excess in $T$ relative to the average. By $z=6$, the
excess in temperature is still present, but at larger distances
($10\lesssim r/{\rm cMpc}\lesssim 30 $); regions further away from the sources have
been more recently photoheated and have had less time to cool.
Gas in the low-density IGM cools adiabatically via the expansion of
the Universe and also through inverse Compton scattering off cosmic
microwave background photons (an important process at these redshifts,
see e.g.  \citealt{puchwein2019}), so the gas closer to the sources
($1\lesssim r/{\rm cMpc}\lesssim 10 $) is instead now cooler than average.  At
$z=5.4$ the picture is similar to $z=6$, but the excess in temperature
occurs still further away from sources ($r \gtrsim 15~{\rm cMpc}$). By
$z=4.2$ the coherent radial fluctuations in temperature have largely
faded, and the temperature more closely correlates with the density
field (i.e., the IGM is relaxing toward a power-law
temperature-density relation, $T=T_{0}\Delta^{\gamma-1}$).  The fact that -- due
to the long cooling timescale for the low density IGM -- relic
fluctuations in temperature persist longer than the fluctuations in
photo-ionisation rate will have consequences for the
\emph{post-reionisation} galaxy--\lya{} transmission correlation
\citep[see also][for related work]{daloisio2018, keating2018,
  molaro2023}. We will discuss this further in \sref{sec:res-phy-rel}.


\section{The galaxy--{\boldmath \lya{}} transmission correlation in Sherwood-Relics}
\label{sec:res}

\subsection{Computing the galaxy--{\boldmath \lya{}} transmission correlation}
\label{sec:res-com-hlt}

We now turn to examine the galaxy--\lya{} transmission correlation
using our mock \lya{} forest spectra that include redshift space
effects. We do this by selecting each halo that satisfies a given mass
constraint (usually $M_{\rm h}\geq 10^{10}~\msolh$) and then calculating
the distance to each pixel in our \lya{} forest sightlines.  We then
use bins of width $\Delta r=147~{\rm ckpc}$ and average the transmission
over many sightlines to obtain $\langle F(r) \rangle$. The galaxy--\lya{}
transmission correlation is then
\begin{equation}
  \label{eq:deltaF}
  \delta_F(r) = \frac{\langle F(r) \rangle}{\bar{F}} - 1,
\end{equation}
where $\bar{F}$ is the global volume-averaged \lya{} transmission at
that redshift, computed from the simulated 
  spectra. Throughout the remainder of this work we therefore also
refer to the galaxy--\lya{} transmission correlation as $\delta_F$.  For the
sake of clarity, we note that $\delta_F<0$ corresponds to less transmission
than the average, and $\delta_F>0$ corresponds to more.

\subsection{Comparison with observational data}
\label{sec:res-com}

\begin{figure}
  \centering
  \includegraphics[width=\linewidth]{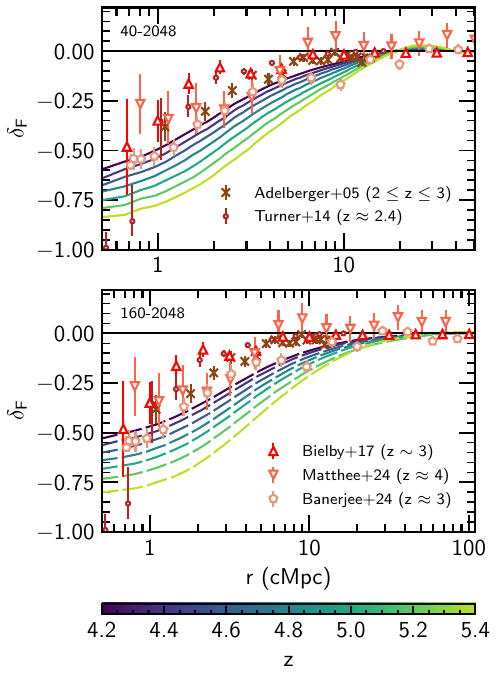}
  \vspace{-0.7cm}
  \caption{The low-redshift ($z\leq 5.4$) galaxy--\lya{} transmission
    correlation ($\delta_F$, see \eref{eq:deltaF}) as a function of
    distance $r$ from haloes with mass $M_{\rm h} \geq 10^{10}~\msolh$ in
    the fiducial 40-2048 (solid, top panel) and 160-2048
    (dashed, bottom panel) simulations. Both models
      are shown every $\Delta z= 0.2$ between $z=4.2$ and $z=5.4$. We also
    show constraints from $2 \leq z \leq 4$ due to \citet{adelberger2005,
      turner2014, bielby2017, matthee2024, banerjee2024}. Note the
    different ranges in $r$ between the top and bottom panels.
    Redshift differences between the data and simulations mean we do
    not expect quantitative agreement. \label{fig:res-F-r-lowz}}
\end{figure}

We begin with a comparison to observational data at $z<4$, well after
reionisation has completed. In \fref{fig:res-F-r-lowz} we compare
these data to $\delta_F$ as a function of distance $r$ from haloes with
$M_{\rm h}\geq 10^{10}~\msolh$ in the 40-2048 and 160-2048 models at
$z \leq 5.4$. This redshift corresponds roughly to the endpoint
  of reionisation in the simulations (i.e., where the volume-averaged
  neutral fraction $\bar{x}_{\rm HI} \le 10^{-3}$, see
  Table~\ref{tab:mod-run}). In both the 40-2048 and 160-2048 runs,
$\delta_F < 0$ close to the haloes, where $\delta_F$ decreases rapidly with
decreasing distance.  Furthermore, at fixed $r$, $\delta_F$ increases with
decreasing redshift.  For the 40-2048 run, $\delta_F$ drops below
$-0.1$ at $r\lesssim 5$--$10{~\rm cMpc}$, depending on redshift, while for
160-2048 the same decrease in $\delta_F$ occurs at
$r\lesssim 10$--$20{~\rm cMpc}$. This decrement in $\delta_F$ over larger distances
in the 160-2048 run is due to the more massive haloes present in the
larger volume, which sit in larger associated overdensities (see
\fref{fig:app-hmf}). For $r>20{~\rm cMpc}$ ($50~{\rm cMpc}$) in the
40-2048 (160-2048) run, the transmission approaches the mean value
(i.e. $\delta_F\rightarrow 0$).

The shape of $\delta_F$ predicted by the models is in qualitative agreement
with observations, and where the redshift of the data points and
simulations are similar ($z\sim 4$) the agreement is reasonable.
However, we do not expect perfect quantitative agreement between our
models and observations, given that the observations are at (often
much) lower redshift (the closest observations to our lowest redshift
snapshot at $z=4.2$ are the \citealt{matthee2024} results, at
$z\approx 4$). In addition, we note again that the lack of a detailed
subgrid model for galaxy formation and feedback (which may be
important for matching observations of $\delta_F \ll 0$ close to galaxies,
see \citealt{meiksin2017, sorini2020}) in Sherwood-Relics means that
we do not expect good agreement close to ($r \lesssim 1{~\rm cMpc}$) haloes.

\begin{figure}
  \centering
  \includegraphics[width=\linewidth]{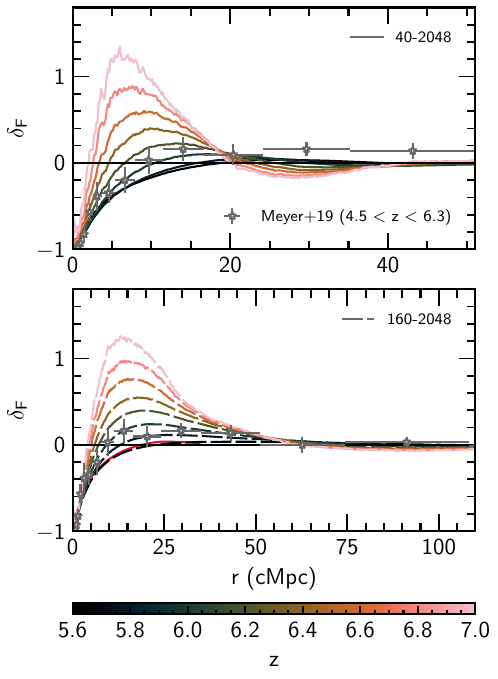}
    \vspace{-0.7cm}
    \caption{The high-redshift ($z > 5.4$) galaxy--\lya{} transmission
      correlation ($\delta_F$, see \eref{eq:deltaF}) as a function of
      distance $r$ from haloes with mass
      $M_{\rm h} \geq 10^{10}~\msolh$ in the fiducial 40-2048 (solid, top
      and bottom panels) and 160-2048 (dashed, bottom
      panel) simulations. We show $\delta_F$ at
        $z=6$ from the 40-2048 model in the bottom panel (solid,
        red) to aid comparison. Both models are shown every
        $\Delta z= 0.2$ between $z=5.6$ and $z=7.0$. Also shown are
      observational constraints due to \citet{meyer2019}, calculated
      using \civ{} absorbers over the redshift range
      $4.5 < z_{\rm \civ} < 6.3$ (with mean redshift
      $\langle z_{\rm \civ}\rangle = 5.18 $) as proxies for galaxies (see
      \sref{sec:mod-rea} for a discussion on the selection criteria
      for haloes). Note the different ranges in $r$ between the top
      and bottom panels.} \label{fig:res-F-r-m19}
\end{figure}

\begin{figure}
  \centering
  \includegraphics[width=\linewidth]{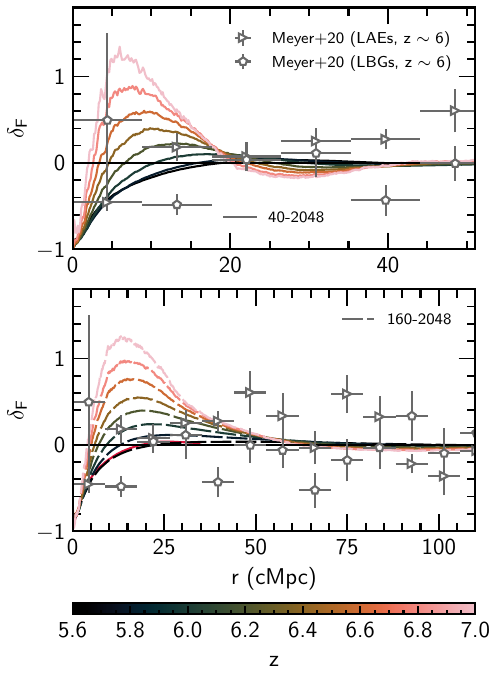}
  \vspace{-0.7cm}
  \caption{As in \fref{fig:res-F-r-m19}, but with constraints
      from \citet{meyer2020}, where $\delta_F$ is computed for Lyman-break
      galaxies (LBGs, triangles) and \lya{} emitters (LAEs, pentagons)
      at $z\sim 6$. \label{fig:res-F-r-m20}}
\end{figure}

\begin{figure}
  \centering
  \includegraphics[width=\linewidth]{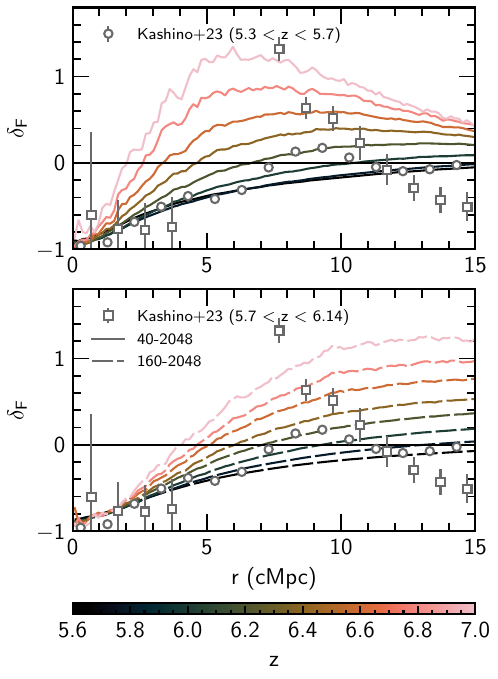}
  \vspace{-0.7cm}
  \caption{As in \fref{fig:res-F-r-m19}, but shown over a
      smaller range in $r$ and with constraints from
      \citet{kashino2023}, where \oiii{} emitters were used in the
      calculation of $\delta_F$. In the interests of readability, the
      higher (lower) redshift \citet{kashino2023} points have been
      offset by 0.2 ($-0.2$) cMpc and we do not show the full extent of
      the higher redshift \citet{kashino2023} data, which extends up
      to $\delta_F \approx 3$ at $r\approx 5 {~\rm cMpc}$. Note that both the top and
      bottom panels now share the same range in
      $r$. \label{fig:res-F-r-k23}}
\end{figure}

\begin{figure}
  \centering
  \includegraphics[width=\linewidth]{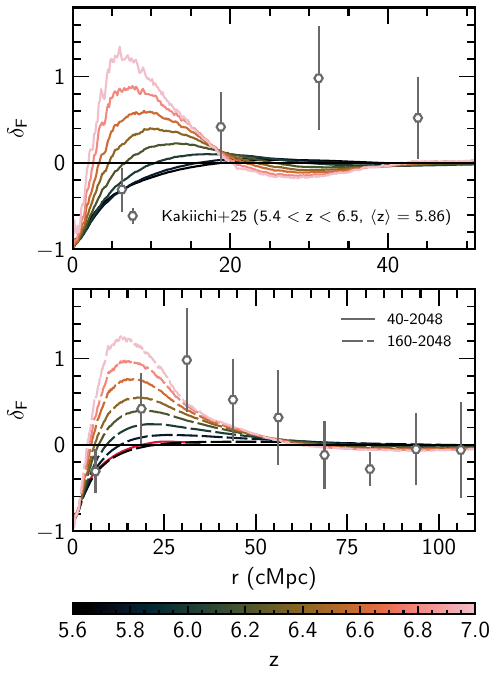}
  \vspace{-0.7cm}
  \caption{As in \fref{fig:res-F-r-m19}, but with constraints from
    \citet{kakiichi2025}, where $\delta_F$ is computed for \oiii{} emitters
    over the redshift range $5.4 < z_{\rm \oiii} < 6.5$ (with mean redshift
    $\langle z_{\rm \oiii}\rangle = 5.86$).}
  \label{fig:app-F-r-k25}
\end{figure}

In Figs. \ref{fig:res-F-r-m19}--\ref{fig:app-F-r-k25} we now focus on the data at redshifts approaching the reionisation era.  We show $\delta_F$ for $z > 5.4$ compared to
  observational constraints from \citet{meyer2019}, \citet{meyer2020}, \citet{kashino2023} and \citet{kakiichi2025}, respectively. Beginning with the
  general features of $\delta_F$ in our models at high $z$, we observe a
similar rapid decrease in $\delta_F$ close to haloes as observed at lower
  $z$, but now $\delta_F > 0$ further away from haloes
($r\gtrsim {\rm few ~cMpc}$). For both the 40-2048 and 160-2048 models, the
amplitude of this excess is larger at higher $z$, when there is very
little transmission in the IGM (i.e. the excess is not driven
by a very large $\langle F(r)\rangle$, but rather the contrast with small
$\bar{F}$). As redshift decreases, the peak of the excess shifts to
larger distances, from $r\approx6{~\rm cMpc}$
($r\approx 12{~\rm cMpc}$) at $z=7$ for 40-2048 (160-2048) to
$r \sim 20 {~\rm cMpc}$ at $z=6$ for both models, although around
$z=6$ the ``peak'' becomes very broad. For $r \gtrsim 20~{\rm cMpc}$
($r\gtrsim 60~{\rm cMpc}$) at $z\geq6$, $\delta_F$ becomes negative again in the
40-2048 (160-2048) model, due to gas in the IGM that has not yet been
highly ionised.

Turning now to comparing the high-$z$ $\delta_F$ in our models to
observational data, we begin with \fref{fig:res-F-r-m19} and the
\citet{meyer2019} constraints. \citet{meyer2019} use the positions of
\civ{} absorbers over the redshift range $4.5 < z_{\rm \civ} < 6.3$
(with mean redshift $\langle z_{\rm \civ} \rangle = 5.18$) as proxies for galaxies
when computing $\delta_F$.  As discussed in \sref{sec:mod-rea}, they argue
that these absorbers approximately correspond to halo masses
$M_{\rm h} \gtrsim 10^{9.8}~\msolh$, although note the observed \civ{} bias
they report is uncertain and is consistent with host halo masses as
large as $M_{\rm h}=10^{12.4}~\msolh$.  For
$r \lesssim 20~{\rm cMpc}$, we find excellent agreement ($< 1\sigma$) between the
\citet{meyer2019} results and our fiducial 40-2048 model, but at $z=6$
as opposed to $\langle z_{\rm \civ} \rangle = 5.18$.  Interestingly, a
qualitatively similar result was also independently noted by
\citet{garaldi2022} using the THESAN simulations, where these authors
found simulation outputs at $z\sim 6.2$ were in better agreement with the
\citet{meyer2019} data.  In addition to the timing discrepancy, for
$r \gtrsim 20~{\rm cMpc}$ we are unable to reproduce the observed
$\delta_F>0$ at any of the redshifts shown. For the 160-2048 model, we also
find excellent agreement at $z=6$ (mostly $<1\sigma$) over the entire range
shown ($110~{\rm cMpc}$).\footnote{That we are able to
    reproduce the observed $\delta_F>0$ for $r \gtrsim 20~{\rm cMpc}$ in the
    160-2048 model, but not the 40-2048 model, suggests that large box
    sizes ($L_{\rm box}>40~\mpch$) are required to model the
    signal at these scales. This presents a challenge to
    theoretical works exploring this effect, given the high resolution
    simultaneously required to correctly model the \lya{} forest
    \citep{bolton2009}.} Again, this redshift, at which the 160-2048
model best agrees with the data, is higher than the average redshift
of \civ{} absorbers in the \citet{meyer2019} dataset. This is perhaps
surprising, given that our simulations are calibrated to reproduce
observations of the mean transmission in the \lya{} forest at
$z\lesssim 6$, but it is equivalent to requiring
$\bar{x}_{\rm HI}\sim 0.1$ at $z=5.2$. For comparison, the distribution
of \lya{} effective optical depths at $z \leq 5.2$ are consistent with a
fully reionised IGM and a homogeneous UV background, suggesting there
is little scope for delaying the end of reionisation below this
redshift \citep{bosman2022}.

We move now to \fref{fig:res-F-r-m20} and the follow-up work of
\citet{meyer2020}, where the cross-correlation is computed for
Lyman-break galaxies (LBGs) and \lya{} emitters (LAEs) at $z\sim6$ as
opposed to \civ{} absorbers at slightly lower redshift. In contrast to
the \citet{meyer2019} result, we find poor agreement with the
\citet{meyer2020} constraints, except for the negative $\delta_F$ at small
($r\lesssim 5~{\rm cMpc}$) scales. Indeed, \citet{meyer2020} note that their
measurement is impacted by the small sample size and noise, and argue
that the two-point cross-correlation function between transmission
spikes (high signal-to-noise pixels with $F>0.02$) and galaxy
positions should provide a more robust measurement. We briefly discuss
this statistic in App.~\ref{sec:app-2pccf}, but focus on the more
widely used $\delta_F$ throughout the main portion of this work.

Next, we turn to \fref{fig:res-F-r-k23} and the comparison with the
\citet{kashino2023} constraints from a single QSO field from the EIGER
survey. \citet{kashino2023} use \oiii-emitting galaxies when
calculating $\delta_F$ and, while we do not directly model \oiii{}
emission, it has been shown that these \oiii{} emitters tend to reside
in haloes with $M_{\rm h}\approx 10^{10.7}~\msolh$ \citep[][see also
discussion in \sref{sec:mod-rea}]{pizzati2024}.
For $z\sim 7$, we reproduce some general features of
the \citet{kashino2023} measurement for $r\lesssim 10 {~\rm cMpc}$, namely
$\delta_F<0$ at small distances ($r\lesssim 4{~\rm cMpc}$) and a strongly-peaked
excess in $\delta_F$ at slightly larger distances
($r \approx 6{~\rm cMpc}$), but are unable to match the extreme value of
$\delta_F\approx 3$ at $r\approx 5\rm\,cMpc$ (not shown in \fref{fig:res-F-r-k23}, see
figure caption) or the behaviour of $\delta_F$ for
$r\gtrsim 10{~\rm cMpc}$.  Our analysis is consistent with the
interpretation of \citet{garaldi2024}, who suggest that the
\citet{kashino2023} result is likely dominated by stochasticity, given
that it focuses on a single QSO sightline. Future results from all six
QSO fields in the EIGER survey could provide interesting new insights
into the nature of the galaxy--\lya{} transmission correlation at high
$z$. In the context of modelling, varying the source model in our
simulations could also prove instructive, particularly for small $r$.

While this work was being completed, new
  results on the \oiii\ emitter--\lya{} transmission correlation were released
  by the {\em JWST} ASPIRE survey \citep{kakiichi2025} using five QSO fields.  In \fref{fig:app-F-r-k25} we show the \citet{kakiichi2025} results compared to our fiducial
  40-2048 and 160-2048 models. For the 40-2048 model at $z=5.8$ --
  approximately the mean redshift of the \oiii{} emitters used in
  \citet{kakiichi2025} -- the simulation is within $1\sigma$ of the
  \citet{kakiichi2025} results for $r \lesssim 20~{\rm cMpc}$. However, the 40-2048 model does
  not match the shape of $\delta_F$ (which peaks around
  $r\approx30~{\rm cMpc}$) at larger $r\gtrsim 20~{\rm cMpc}$. For the 160-2048
  model at $z=5.8$, the agreement is much better and the simulation is within $1.5 \sigma$ of the
  \citet{kakiichi2025} results over the entire range of $r$ shown, although the peak in $\delta_F$ occurs at a smaller
  $r\approx20~{\rm cMpc}$ in the 160-2048 model than in the observational
  data. As we will discuss in Sec.~\ref{sec:res-imp}, selecting a larger
  minimum halo mass for the haloes used in the calculation of
  $\delta_F$ may improve the agreement further
  (cf. \fref{fig:res-F-r-mbin}).  A box size larger than $160~h^{-1}\rm\,cMpc$ may also be required to fully capture the largest ionised bubbles and the ionising sources clustering around massive haloes.

\subsection{The galaxy--Ly~{\boldmath $\alpha$} transmission correlation and the local mean free path around haloes}
\label{sec:res-gal}

Finally, we explore the connection between the location of the peak of
$\delta_F$ in Figs. \ref{fig:res-F-r-m19}--\ref{fig:app-F-r-k25} and the local mean free path of Lyman-limit
photons around galaxies, as highlighted recently by
\citet{garaldi2024}. In this case, we calculate $\delta_F$ without
including the effect of peculiar velocities to enable a fair
comparison with the real-space \lya{} transmission defined in
\eref{eq:sim-Freal}. We compute the distance at which $\delta_F$ is
largest, \rdfm{}, by first smoothing $\delta_F$ with a moving window of
eight times the bin width ($\Delta r=1180~{\rm ckpc}$) to reduce noise (for
comparison, \citealt{garaldi2024} use a window of $2950~{\rm ckpc}$).

We show \rdfm{} as a solid circle in \fref{fig:sim-proj} and
\fref{fig:sim-proj-qty}. At all redshifts, \rdfm{} occurs well away
from the high-density peak associated with the haloes. At $z=7$ and
$z=6$, \rdfm{} sits inside the region where the radiation field is
enhanced relative to the background, and the hydrogen is
correspondingly more ionised. At $z=5.4$ however, after the radiation
field becomes close to homogeneous, the (very small) peak that remains
for the excess transmission approximately coincides with the residual
excess in the gas temperature. The behaviour of the halo mean free
path, $\Lambda_{\rm mfp}$, measured from Sherwood-Relics by
\citet{feron2024}\footnote{\citet{feron2024} measure
  $\Lambda_{\rm mfp}$ from sightlines originating at the virial radius of
  haloes (see their sec.~4 for details).} (dashed circles) is
qualitatively similar to that of \rdfm{}.  There will be small
differences due, in part, to the slightly different sample of haloes
used to calculate $\Lambda_{\rm mfp}$ in \citet{feron2024} (see also
\sref{sec:res-imp}).  Nevertheless, as first noted recently by
\citet{garaldi2024}, this suggests the typical scale for the excess
transmission in the galaxy--\lya{} transmission correlation is related
to the local mean free path for ionising photons around
galaxies. Prior to the end of reionisation, this quantity will be
larger than the mean free path in the average IGM, but following
reionisation it will be smaller (see e.g., fig. 8 in
\citealt{feron2024} and the related discussion, as well as
\citealt{fan2025}).


\section{The detailed physical origin of the galaxy--{\boldmath \lya{}} correlation}
\label{sec:res-phy-main}

\subsection{Degeneracy between halo mass and neutral fraction}
\label{sec:res-imp}

\begin{figure}
  \centering
  \includegraphics[width=\columnwidth]{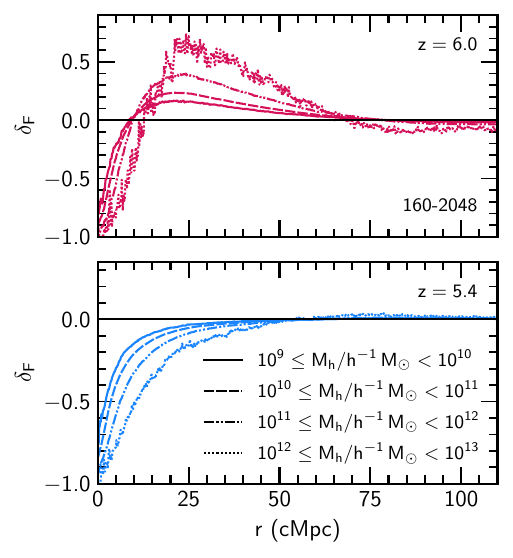}
  \vspace{-0.8cm}
  \caption{Impact of halo mass used in the calculation of
    the galaxy-\lya{} transmission fluctuation correlation in the
    160-2048 run at $z=6.0$ (top panel, pink) and $z=5.4$
    (bottom panel, blue). We show the correlation
    between our simulated spectra and haloes in bins
    spanning four decades in mass:
    $10^{9}\leq M_{\rm h}/\msolh<10^{10}$ (solid),
    $10^{10}\leq M_{\rm h}/\msolh<10^{11}$ (dashed),
    $10^{11}\leq M_{\rm h}/\msolh<10^{12}$ (dash-dot-dotted), and
    $10^{12}\leq M_{\rm h}/\msolh<10^{13}$
    (dotted). \label{fig:res-F-r-mbin}}
\end{figure}

We now turn to more closely examine the various factors that influence
the galaxy--\lya{} correlation in our simulations. First, we examine
the galaxy--\lya{} transmission correlation computed using
  only haloes in a given mass range in \fref{fig:res-F-r-mbin}.  We
consider the results from the 160-2048 model at $z=6$ and $z=5.4$,
although note that the general trends are the similar in the 40-2048
box. Here we use the 160-2048 box instead of the fiducial,
higher-resolution 40-2048 box because it contains haloes with larger
masses, $M_{\rm h} > 10^{12}~\msolh$ while also still resolving the
minimum halo mass $M_{\rm h} = 10^{9}~\msolh$ that hosts ionising
sources in our simulations (see e.g., \fref{fig:app-hmf}).  We compute
$\delta_F$ for haloes spanning four decades in mass, although note the
lowest mass bin ($10^{9}\leq M_{\rm h}/\msolh<10^{10}$) is shown only for
comparison here and is not used throughout the rest of this work.  For
mass bins with $M_{\rm h}\ge10^{10}~\msolh$ we use all haloes when
calculating $\delta_F$ and, for computational efficiency, we draw a
representative sample of $10^4$ haloes for those with
$10^{9}\leq M_{\rm h}/\msolh<10^{10}$. At $z=6$ ($z=5.4$) the largest
halo has a mass of $4.5\times 10^{12}~\msolh$ ($6.9\times10^{12}~\msolh$).

\begin{figure}
  \centering
  \includegraphics[width=\columnwidth]{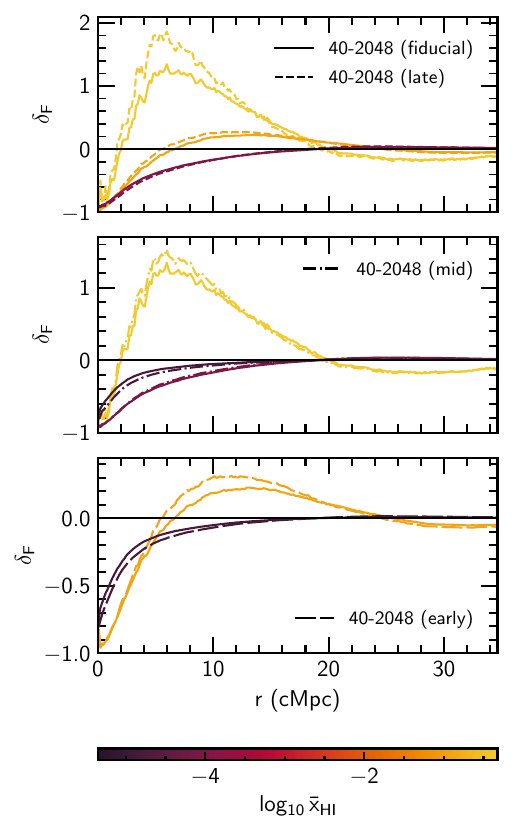}
  \vspace{-0.7cm}
  \caption{Comparison of different reionisation histories to the
    fiducial 40-2048 run at approximately equal volume weighed IGM
    neutral hydrogen fractions, $\bar{x}_{\rm HI}$.  Shown are the
    late (top panel, dashed), mid (middle panel dot-dashed) and early
    (bottom panel, long dashed) reionisation models along with the
    fiducial model (solid, with the same curves are shown in all
    panels). The colour indicates $\log_{10}\bar{x}_{\rm HI}$.  In
    each case we select the closest match in $\bar{x}_{\rm HI}$
    between the fiducial model values and each of the three
      reionisation models measured at $z=7, 6$ and $5.4$. We do not
      show $\delta_F$ at $z=5.4$ for the early model, as the fiduical model
      does not extend to low enough redshift to match the neutral
      hydrogen fraction. Note the neutral fraction decreases with
    decreasing redshift, so in each panel the curves with larger
    (smaller) $\bar{x}_{\rm HI}$ correspond to higher (lower) redshift
    outputs for each model.}
  \label{fig:res-F-r-zr}
\end{figure}

In the lower panel of \fref{fig:res-F-r-mbin} at $z=5.4$ (close to the
end of reionisation), for $r\lesssim 50~{\rm cMpc}$ we find that the radial
extent of the region with negative $\delta_F$ becomes larger as the halo
mass used to calculate $\delta_F$ increases. This is because the
lower-mass haloes are associated with smaller overdensities where the
clustering and infall of neutral hydrogen is weaker. In
photoionisation equilibrium, $\Delta \propto \xhi$, and in the absence of a
strong proximity effect or feedback we therefore expect the gas
approaching the lower-mass haloes to have smaller $\xhi$ (i.e., are
more highly ionised) and larger transmission (i.e., a smaller region
of negative $\delta_F$).

The results displayed in the upper panel of \fref{fig:res-F-r-mbin} at
$z=6$ are more complicated, however. Due to proximity to the ionising
sources, the gas within $r\lesssim 5{~\rm cMpc}$ is reionised earlier than
gas further away \citep[see e.g., fig.~1
in][]{puchwein2023}. Therefore, by $z=6$, this gas is already in
photoionisation equilibrium and the behaviour observed at $z=5.4$ is
replicated.  However, further away from haloes
($10 \lesssim r/{\rm cMpc}\lesssim 70$), it is now the more massive haloes which
have larger, positive $\delta_F$ (i.e., excess transmission relative to the
mean). This is because, in the Sherwood-Relics model, the ionising
emissivity of a source, $\dot{N}_\gamma$, is proportional to halo mass
\citep{kulkarni2019}; when the IGM is optically thin to ionising
photons the photoionisation rate in the low-density IGM (i.e., far away
from haloes where local density effects become unimportant) is always
larger around more massive haloes due to source clustering.  We
therefore speculate that simulations which use a different source
luminosity assignment, such as a `Democratic' model where
$\dot{N}_\gamma$ is independent of $M_{\rm h}$ \citep{cain2023,
  asthana2024}, may change these results.  Temperature fluctuations
also play a role, as gas that is ionised (and therefore heated)
earlier has more time to cool than gas which is ionised later. At a
given density, colder gas is more neutral (since the recombination
coefficient scales as $\alpha \propto T^{-0.72}$ for gas
$T\sim 10^4~{\rm K}$) and so the (relatively) cooler gas near to haloes
will also exhibit reduced transmission compared to the hotter gas
farther away.

In \fref{fig:res-F-r-zr}, we instead compare $\delta_F$ for the three
additional reionisation models listed in \tref{tab:mod-run} to the
fiducial model (solid curves in each panel). The redshift evolution of
the galaxy--\lya{} transmission correlation is different for each
reionisation model.  Hence, rather than compare these models at fixed
redshift, in \fref{fig:res-F-r-zr} we instead compare the models
to the fiducial model at approximately the same volume
averaged neutral hydrogen fraction. We do this by finding the
  closest match in $\bar{x}_{\rm HI}$ between the fiducial model and
  each reionisation model. In Table~\ref{tab:zr-xhi} we list the
  redshifts and corresponding neutral hydrogen fractions for the
  matched outputs. For the different reionisation models, we use
  snapshots at $z=7,\,6$ and $5.4$, except for the early model where
  we only use snapshots at $z=7$ and $6$. This is because the fiducial
  model was not run to low enough redshift to match the neutral
  hydrogen fraction at $z=5.4$ in the early model. 
  
  We find that each
model shows similar behaviour at fixed neutral fraction, suggesting
that the shape of $\delta_F$ is determined primarily by the neutral
fraction in the IGM. This is consistent with the results of
\citet{garaldi2022}, who find that comparing $\delta_F$ at a given
$\bar{x}_{\rm HI}$ reduces the scatter between models with different
reionisation histories.  We note, however, that this will be at least
partly degenerate with the selected halo mass; a larger excess transmission is
observed for either a larger volume averaged neutral hydrogen
fraction, or for a larger assumed halo mass.  In both cases, the IGM
around the haloes hosting sources will be more highly ionised than the
average IGM.  It may therefore be challenging to use the galaxy--\lya{}
transmission correlation as simple proxy for either the volume
averaged neutral fraction or halo mass. Nevertheless, this degeneracy
also hints at a plausible route to (at least partly) resolving the
redshift / \hi{} fraction mismatch between our simulations and the
\citet{meyer2019} measurements discussed in \sref{sec:res-com}.  The
cross-correlation with larger halo masses should facilitate better
agreement with the \citet{meyer2019} measurements at smaller volume
averaged neutral fractions, yielding better agreement with the
observed \lya{} forest effective optical depth distribution at
$z\sim 5.2$.  Note again the possibility of a larger minimum host halo
mass than our fiducial assumption of $M_{\rm h}= 10^{10}~\msolh$ is
not ruled out by the rather uncertain \civ{} absorber bias reported by
\citet{meyer2019}.  Carefully investigating this will require
simulations with larger volumes that sample a representative number of
high mass haloes, however; we plan to investigate this further in
future work.

\begin{table}
  \centering
  \begin{tabular}{lcccc}
    \hline
    \multirow{1}{*}{Reionisation model} 
                                        &$z$ & $\bar{x}_{\rm HI}$& $z_{\rm match}$& $\bar{x}_{\rm HI,match}$ \\
    \hline
    \multirow{3}{*}{late ($z_{\rm r}=5.3$)} & $7.0$ & $0.47$ & $7.0$ & $0.41$ \\
                       & $6.0$ & $0.14$ & $6.2$ & $0.12$ \\
                       & $5.4$ & $10^{-2.48}$ & $5.6$ & $10^{-4.08}$ \\
    \hline
    \multirow{3}{*}{mid ($z_{\rm r} = 6.0$)} & $7.0$ & $0.44$ & $7.0$ & $0.41$ \\ 
                       & $6.0$ & $10^{-2.62}$ & $5.6$ & $10^{-4.08}$ \\
                       & $5.4$ & $10^{-4.90}$ & $4.2$ & $10^{-4.86}$ \\
    \hline
    \multirow{2}{*}{early ($z_{\rm r} = 6.6$)} & $7.0$ & $0.16$ & $6.2$ & $0.12$ \\
                       & $6.0$ & $10^{-5.11}$ & $4.2$ & $10^{-4.86}$ \\
    \hline
  \end{tabular}
  \caption{Redshifts and neutral fractions for the $\delta_F$ shown
      in \fref{fig:res-F-r-zr}. Listed in the columns are: the model name and reionisation redshift, $z_{\rm r}$ (see also Table~\ref{tab:mod-run}), the corresponding volume weighted neutral hydrogen fraction, $\bar{x}_{\rm HI}$, at redshift $z$ in each of these models, and the 
      redshift, $z_{\rm match}$, and volume weighted neutral hydrogen fraction, $\bar{x}_{\rm HI,match}$, in our fiducial reionisation model ($z_{\rm r}=5.7$) that
      most closely matches  $\bar{x}_{\rm HI}$. We do not
      include the early model at $z=5.4$, as the fiducial model does
      not extend to low enough redshift to provide a good match in
      $\bar{x}_{\rm HI}$.}
  \label{tab:zr-xhi}
\end{table}

\subsection{The effect of inhomogeneous reionisation}
\label{sec:res-phy}

We may further tease apart the physical processes driving the specific
form of $\delta_F$ presented in this work by performing a rescaling of our
simulation data, each time isolating a different physical effect.  As
before, all calculations are performed on the fiducial 40-2048 model,
including all haloes with $M_{\rm h} \geq 10^{10}~\msolh$ and, unless
otherwise stated, \emph{without} including the effects of peculiar
velocities, to simplify our interpretation of the results.

\begin{figure}
  \centering
  \includegraphics[width=\columnwidth]{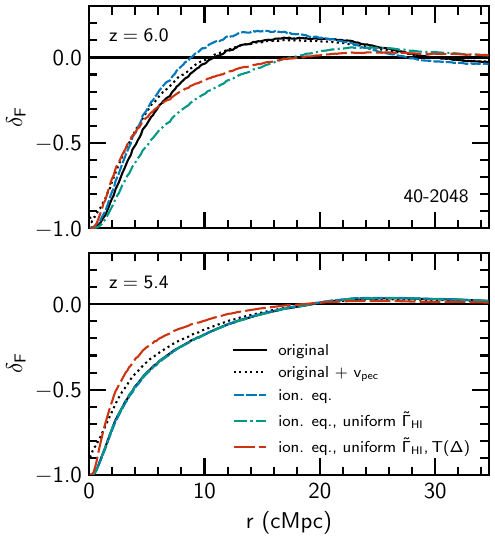}
  \vspace{-0.7cm}
  \caption{Galaxy--\lya{} transmission correlation calculated
    from the 40-2048 run (`original', black solid), after: recomputing
    the ionisation fractions of hydrogen and helium assuming
    ionisation equilibrium (`ion.\ eq.', blue short dashed); setting
    the photoionisation rate to be uniform (`\ieug{}', green
    dot-dashed); and after removing temperature fluctuations by
    setting $T\propto \Delta^{\gamma -1}$ and rescaling the neutral fraction
    appropriately (`\ieugtd{}', red long dashed). Also shown is the
    original case including peculiar velocities
    (`original~+~$v_{\rm pec}$', black dotted) -- all other curves do
    not include the effect of peculiar velocities on \lya{}
    transmission. We show results at $z=6$ (top panel) and
    $z=5.4$ (bottom panel). Note that at $z=5.4$ the
    curves lie directly on top of one another, except for
    the `original~+~$v_{\rm pec}$' and `\ieugtd{}' cases.}
  \label{fig:res-F-r-toy}
\end{figure}

In \fref{fig:res-F-r-toy} we show the form of $\delta_F$ when including
different physical effects at $z=6.0$ and $z=5.4$. Starting with the
dotted black curve (`original~+~$v_{\rm pec}$'), when compared to the
solid black curve (`original'), this shows the effect of including
peculiar velocities $v_{\rm pec}$ in the optical depth calculation.
Note that the dotted black curves are the same as those shown in
Figs. \ref{fig:res-F-r-m19}, \ref{fig:res-F-r-m20} and
\ref{fig:res-F-r-k23}. We find that including $v_{\rm pec}$ makes only
a modest difference to the shape of $\delta_F$ at both redshifts shown, and
so plays a minimal role in setting the shape of the galaxy--\lya{}
transmission correlation toward the tail-end of reionisation.
  
Next we turn to the blue short-dashed curve in \fref{fig:res-F-r-toy}
(`ion.\ eq.'), which shows $\delta_F$ calculated from sightlines where the
ionic abundances have been recalculated assuming the gas is in
ionisation equilibrium with the spatially inhomogeneous UV background used in the simulation. Contrasting the blue short-dashed curves with the solid
black curves isolates the impact of non-equilibrium effects on
$\delta_F$. At $z=5.4$, non-equilibrium effects have no impact on
$\delta_F$. This is not surprising, given that reionisation has ended and
so the gas is already largely in ionisation equilibrium. In contrast,
non-equilibrium effects play a role in modifying the shape of
$\delta_F$ at $z=6$, though qualitatively the picture remains the same
(i.e. the excess in $\delta_F$ remains after enforcing ionisation
equilibrium). Again, it is not surprising that non-equilibrium effects
are important during reionisation, because not all of the gas will be in
ionisation equilibrium. \citet{zhu2024} also explore the impact of assuming
ionisation equilibrium on $n_{\rm HI}$, finding that it leads to
deviations at the few~per~cent level.

The green dot-dashed curve in \fref{fig:res-F-r-toy} (`\ieug{}') also
assumes ionisation equilibrium, but this time we do not use the
inhomogeneous UV background from the simulation. Instead, we assume that the UV background
is spatially uniform with photoionisation rate
$\tilde{\Gamma}_{\rm HI}$. Comparing the green dot-dashed and blue
short-dashed curves therefore highlights the importance of UV background
fluctuations due to patchy reionisation. For
$\tilde{\Gamma}_{\rm HI}$ we use the time-varying, average photoionisation
rate measured at a distance of $1.5~{\rm cMpc}$ away from all haloes
with $M_{\rm h} \geq 10^{10}~\msolh$ in the fiducial 40-2048
simulation. The exact value of $\tilde{\Gamma}_{\rm HI}$, or the distance
at which it is measured, is of little import -- the key point is that
$\tilde{\Gamma}_{\rm HI}$ must be large enough to induce small enough
equilibrium values of $x_{\rm HI}$ to allow for \lya{} transmission; at $z=6$ ($z=5.4$)
$\tilde{\Gamma}_{\rm HI}=2.0\times10^{-13}~{\rm s}^{-1}$
($3.2\times10^{-13}~{\rm s}^{-1}$). At $z=6$, the effect of the uniform UVB
is striking. The large excess in $\delta_F$ around
$10\lesssim r/{\rm cMpc} \lesssim 25$ is no longer present. Instead, the
$z=6$ green dot-dashed curve most closely resembles its
post-reionisation counterpart at $z=5.4$, where the shape of
$\delta_F$ is characterised by $\delta_F< 0$ for
$r\lesssim 20~{\rm cMpc}$. From this, we infer that the excess in
$\delta_F$ present in the original simulation is a direct consequence of
the inhomogeneous nature of reionisation, i.e. the excess is driven by
local ionisation enhancement due to ionising sources, or clustered
sources. This also highlights the importance of contrast with
$\delta_F$ -- if $\bar{F}$ is small (e.g., due to large neutral fractions in
the average IGM) then even a small increase in transmission can lead
to a large excess in $\delta_F$. At $z=5.4$, enforcing a uniform UV background makes
no difference to $\delta_F$, as the radiation field is already almost entirely uniform
(cf. the fluctuations in $\Gamma_{\rm HI}$ in \fref{fig:sim-proj-qty}).

Finally, we turn to the red long-dashed curve in
\fref{fig:res-F-r-toy} (`\ieugtd{}'), which examines the effect of
temperature fluctuations on $\delta_F$. To do this, we take the sightlines
with a uniform UV background (i.e. the sightlines used to produce the green
dot-dashed curve) and fit the relation
\begin{equation}
  \label{eq:TD}
  T=T_0\Delta^{\gamma-1}
\end{equation}
to the low-density ($\Delta < 10$) ionised ($x_{\rm HI}<10^{-3}$) gas at
each $z$. To remove any effect due to the redshift evolution of $T_0$
and $\gamma$, we fix their value to the average of the $z=6$ and
$z=5.4$ values, taking $T_0=1.14\times10^4~{\rm K}$ and $\gamma=1.15$. These
parameter values are consistent with the \citet{gaikwad2020}
constraints, which cover almost exactly the same redshift range. We
then recalculate the temperature in every pixel using \eref{eq:TD} and
the fitted parameters $T_0$ and $\gamma$, thus removing spatial
fluctuations in temperature due to inhomogeneous reionisation. The
corresponding original neutral fraction $x_{\rm HI, orig}$ in each
pixel is then rescaled, using the original temperature $T_{\rm orig}$
and assuming photoionisation equilibrium, according to
\begin{equation}
  \label{eq:xHIre}
  x_{{\rm HI}, T(\Delta)} = x_{\rm HI,orig}\frac{\alpha_{A}[T(\Delta)]}{\alpha_A(T_{\rm orig})},
\end{equation}
where $\alpha_A$ is the case-A recombination rate calculated using the fit
in \citet{verner1996}. At both redshifts, we see that removing
temperature fluctuations (red long dashed curves) makes $\delta_F$ larger (less negative) closer to haloes
($r\lesssim 15~{\rm cMpc}$) compared to the case with a uniform UV background and temperature
fluctuations (green dot-dashed curves). This is because, in the patchy reionisation scenario,
gas close to haloes is heated earlier than gas farther away, and
therefore has more time to cool (cf. right column of
\fref{fig:sim-proj-qty}). Setting $T\propto \Delta^{\gamma - 1}$ and performing the
appropriate rescaling of $x_{\rm HI}$ removes these fluctuations and
thus reduces the contrast between small $r$ (where gas was initially
slightly colder) and large $r$ (where gas was initially slightly
warmer), leading to the relative increase in $\delta_F$ at small $r$.  In the lower panel of \fref{fig:res-F-r-toy} we observe this effect persists after fluctuations in the UV background have faded away \citep[see also][for the importance of this effect in the context of the connection between \lya{} opacity and galaxy density]{christenson2023,gangolli2024,garaldi2024a}.

\subsection{Relic post-reionisation temperature fluctuations}
\label{sec:res-phy-rel}
Finally, we explore in more detail how temperature fluctuations affect the
post-reionisation evolution of $\delta_F$ at $4.2\leq z \leq5.2$.  To isolate the temperature fluctuations we use the average $T_0$ and $\gamma$ in the models over this redshift range to once again remove any effect due to the redshift evolution of these parameters. We use
$T_0=1.14\times10^4~{\rm K}$ and $\gamma=1.25$; this $\gamma$ is consistent
with the \citet{boera2019} results but $T_0$ is somewhat larger
than that reported in that study.

\begin{figure}
  \centering
  \includegraphics[width=\linewidth]{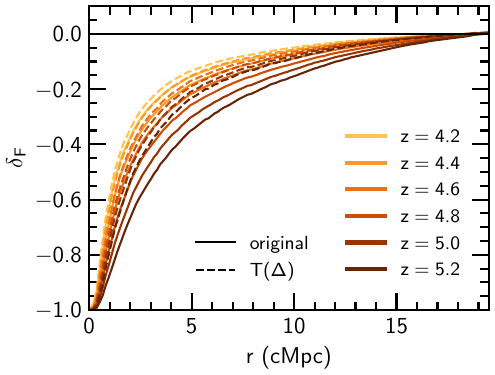}
  \vspace{-0.7cm}
  \caption{Post-reionisation galaxy--\lya{} transmission
    correlation from the original slightlines (`original', solid) and
    where temperatures have been recalculated according to
    $T \propto \Delta^{\gamma - 1}$  to remove spatial fluctuations in the gas temperature and $x_{\rm HI}$ has been appropriately rescaled
    (`$T(\Delta)$', dashed), shown from $z=5.2$ (dark brown) to $z=4.2$
    (gold).   
    \label{fig:prT}}
\end{figure}

In \fref{fig:prT}, we show the post-reionisation evolution of
$\delta_F$ with (solid curves) and without (dashed curves) the post
reionisation temperature fluctuations, focusing on the region close to
the haloes where excess absorption (negative $\delta_F$) occurs.  The
redshift evolution of the excess absorption in the original case is
much stronger than in the rescaled $T(\Delta)$ case, where the evolution is
now instead driven almost entirely by the redshift evolution of the
mean transmission. Therefore, at these post-reionisation redshifts, we
expect that relic temperature fluctuations due to inhomogeneous
reionisation should play a significant role in setting the
galaxy--\lya{} transmission correlation in proximity to the haloes
hosting ionising sources. As redshift decreases, the role of
temperature fluctuations diminishes (i.e. the original case approaches
the $T(\Delta)$ case). This is in line with predictions from previous
studies which have found that the effect of relic temperature
fluctuations (due to inhomogeneous reionisation) on e.g., the 1D
\lya{} forest power spectrum decrease with decreasing redshift
\citep[e.g.,][]{molaro2022}.  Recent work by \citet{matthee2024} has
begun to probe the excess small-scale \lya{} absorption around
galaxies in the post-reionisation epoch ($z\approx 4$). Extending this work
to study the redshift evolution of this excess absorption could therefore
provide insights into the nature of post-reionisation relic
temperature fluctuations, and consequently on the timing of
reionisation.

\section{Conclusions}
\label{sec:con}
In this work, we have explored the cross-correlation between
galaxies and \lya{} transmission in the intergalactic medium, using
the Sherwood-Relics simulations \citep{puchwein2023}. The
Sherwood-Relics suite uses a novel hybrid radiative transfer approach,
which self-consistently captures the hydrodynamic response of gas to
inhomogeneous reionisation, and is designed to accurately model the
\lya{} forest. Our main findings are as follows:

\begin{itemize}
\item We find quantitative agreement between the galaxy--\lya{}
  transmission correlation predicted by Sherwood-Relics model and
  \citet{meyer2019} (who use \civ{} absorbers as a proxy for galaxies)
  if the galaxies are hosted in haloes with masses
  $M_{\rm h}\geq 10^{10}~\msolh$, although only if there is a
  redshift mismatch between simulation and observation; the agreement
  occurs for higher redshifts in our models, $z=6$, than the average
  redshift of $z=5.2$ in the \citet{meyer2019} measurements \citep[see
  also][for a similar result]{garaldi2022}. However, this redshift
  mismatch is equivalent to requiring $\bar{x}_{\rm HI}\sim 0.1$ at
  $z\simeq 5.2$, which is inconsistent with the observed \lya{} forest
  effective optical depth distribution at $z=5.2$ \citep{bosman2022}.
  We suggest this apparent tension may be partly resolved if the
  minimum \civ{} absorber host halo mass is instead larger than
  $M_{\rm h}=10^{10}~\msolh$, which is still consistent with the
  \civ{} bias reported by \citet{meyer2019}.  We find poorer
  agreement with \citet{meyer2020} and \citet{kashino2023}, although these measurements are more heavily impacted by limited sample size and noise.\\

\item We find reasonable quantitative agreement ($< 1.5 \sigma$) between the recent \oiii\ emitter -- \lya\ transmission correlation reported by \citet{kakiichi2025} at $z=5.8$ from the {\em JWST} ASPIRE survey, but only when using our larger $160^{3}~h^{-1}\rm\,cMpc^{3}$ simulation volume.  The peak in $\delta_F$ occurs at a smaller $r\approx20~{\rm cMpc}$ than in the observational data when selecting haloes with $M_{\rm h}>10^{10}~\msolh$.  Selecting a larger minimum halo mass and/or a larger simulation box size that captures the ionising source clustering around massive haloes may improve agreement further.  By contrast, our smaller simulation volume of $40^{3}~h^{-3}\rm\,cMpc^{3}$ fails to match the \citet{kakiichi2025} data at $r>20~\rm\,cMpc$.\\
  
\item We also find qualitative agreement between the shape of the
  galaxy--\lya{} transmission correlation in the Sherwood-Relics models
  and a variety of measurements of the galaxy--\lya{} transmission correlation at $z\leq 4$.  However, we are unable to perform a direct comparison as our simulation outputs do not extend below $z=4$. \\

\item The galaxy--\lya{} transmission correlation is sensitive to the
  volume averaged neutral fraction of the IGM, and is largely
  insensitive to different reionisation histories after accounting for
  differences in the neutral fraction (at least for the source model
  used in Sherwood-Relics, which assumes ionising luminosity is
  proportional to halo mass). However, a partial degeneracy arises
  from the assumed host halo masses for the galaxies used in the
  calculation of the correlation.  This degeneracy means it may
  require an independent measure of the host halo masses to use the
  galaxy--\lya{} transmission correlation
  as a simple proxy for the IGM neutral fraction.\\

\item During reionisation, excess transmission around galaxies is
  driven largely by the enhanced ionising radiation field due to
  clustered ionising sources.  Spatial fluctuations in the IGM also
  play a small role, due to the hotter gas present in recently
  reionised regions.  Non-equilibrium effects only have a modest
  impact on the galaxy--\lya{} transmission correlation (see e.g.,
  \citealt{zhu2024}). The distance where the peak excess transmission
  in the cross-correlation occurs is closely related to the the mean
  free path of Lyman-limit photons around the galaxy host haloes (see
  also
  \citealt{feron2024, fan2025, garaldi2024}). \\
  
\item After reionisation, for $r\lesssim 20~{\rm cMpc}$, relic spatial
  fluctuations in the IGM temperature affect the shape of the
  galaxy--\lya{} transmission correlation. As redshift decreases, the
  strength of this effect decreases, and it should fade by $z\approx4$
  \citep[e.g.,][]{molaro2022}. The physical explanation is that gas
  close to haloes is reionised earlier, and thus has more time to
  cool, in contrast to gas at larger distances which is reionised
  later and is comparatively hotter. Observations are beginning to
  probe the galaxy--\lya{} transmission correlation at
  $4\lesssim z \lesssim 5$ \citep{matthee2024}.  Constraining the redshift evolution
  of the correlation over this period, at scales of a
  $\sim\text{few~cMpc}$, could provide complementary insight into the
  timing of reionisation via the signature of post-reionisation
  temperature fluctuations.
\end{itemize}

In summary, the galaxy--\lya{} transmission correlation is an exciting
probe of the connection between high-redshift galaxies and the
intergalactic medium, with the potential to offer new insights into
the timing and main drivers of reionisation. At present, the
high-redshift picture is limited by a paucity of data, but current --
e.g., the \emph{JWST} surveys EIGER \citep{kashino2023} and ASPIRE
\citep{jin2024,kakiichi2025} -- and future observing programmes will shed
more light on this promising probe of the tail end of
reionisation. Additional work on forward modelling of simulations to
better reflect the observations (e.g. for the galaxy selection
criteria) could prove fruitful \citep{garaldi2024a}. We also suggest
that further modelling that explores the impact of different source
models (particularly where the relationship between halo mass and
ionising emissivity is varied, e.g., \citealt{cain2023, asthana2024},
but see also \citealt{garaldi2022, garaldi2024}) will be especially
useful in linking the properties of ionising sources with the
intergalactic medium.  Finally, larger volume simulations
($>160^{3}h^{-3}\,\rm cMpc^{3}$) with sufficient mass resolution to
resolve the \lya{} forest transmission at $z>5$ will also be valuable
for exploring the galaxy--\lya{} correlation on large scales and for
galaxy host halo masses $M_{\rm h}>10^{12}~\msolh$.

\section*{Acknowledgements}
We thank the anonymous referee for a constructive report, and Enrico
Garaldi, Koki Kakiichi, Daichi Kashino and Romain Meyer for insightful
discussions and sharing data. The simulations used in this work were
performed using the Joliot Curie supercomputer at the Tr{\'e}s Grand
Centre de Calcul (TGCC) and the Cambridge Service for Data Driven
Discovery (CSD3), part of which is operated by the University of
Cambridge Research Computing on behalf of the STFC DiRAC HPC Facility
(www.dirac.ac.uk).  We acknowledge the Partnership for Advanced
Computing in Europe (PRACE) for awarding us time on Joliot Curie in
the 16th call. The DiRAC component of CSD3 was funded by BEIS capital
funding via STFC capital grants ST/P002307/1 and ST/R002452/1 and STFC
operations grant ST/R00689X/1.  This work also used the DiRAC@Durham
facility managed by the Institute for Computational Cosmology on
behalf of the STFC DiRAC HPC Facility. The equipment was funded by
BEIS capital funding via STFC capital grants ST/P002293/1 and
ST/R002371/1, Durham University and STFC operations grant
ST/R000832/1. DiRAC is part of the National e-Infrastructure.  LC and
JSB are supported by STFC consolidated grant ST/X000982/1.  Support by
ERC Advanced Grant 320596 `The Emergence of Structure During the Epoch
of Reionization' is gratefully acknowledged. MGH has been supported by
STFC consolidated grants ST/N000927/1 and ST/S000623/1.  We thank
Volker Springel for making P-Gadget-3 available. We also thank
Dominique Aubert for sharing the ATON code, and Philip Parry for
technical support. This work made use of the following open-source
software packages: {\sc cmasher} \citep{vandervelden2020}; {\sc
  matplotlib} \citep{hunter2007}; {\sc numpy} \citep{harris2020}; {\sc
  open-cv} \citep{opencv_library}; and {\sc scipy}
\citep{virtanen2020}.



\section*{Data Availability}
All data and analysis code used in this work are available from the
first author on reasonable request.  Further guidance on accessing the
Sherwood-Relics simulation data may also be found at
\url{https://www.nottingham.ac.uk/astronomy/sherwood-relics/}.



\bibliographystyle{mnras}
\bibliography{refs} 




\appendix

\section{Resolution convergence}
\label{sec:app-con}

\begin{figure}
  \centering
  \includegraphics[width=\columnwidth]{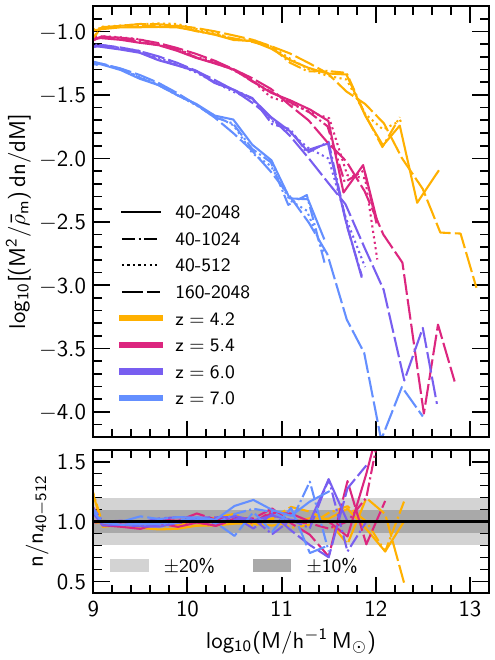}
  \vspace{-0.7cm}
  \caption{Halo multiplicity functions (top panel) for the 40-2048 (solid),
    40-1024 (dot-dashed), 40-512 (dotted), and 160-2048 (long-dashed)
    runs, at $z=7.0$ (blue), $z=6.0$ (purple), $z=5.4$ (dark pink) and
    $z=4.2$ (gold). We also show the ratio of 40-2048, 40-1024 and 160-2048
    to 40-512 at each redshift (bottom panel), indicating the 20 per cent and 10 per
    cent scatter with light and dark grey bands,
    respectively.}
  \label{fig:app-hmf}
\end{figure}

\fref{fig:app-hmf} shows the halo multiplicity function
\citep[e.g.,][]{jenkins2001} 
\begin{equation}
  \label{eq:hmf-mult-sim}
  f(\sigma, z)\frac{\ud \ln \sigma^{-1}}{\ud \ln M} = \frac{M^2}{\bar{\rho}_{\rm m}(z)}\frac{\ud n(M, z)}{\ud M},
\end{equation}
where $\sigma^2$ is the variance of the linear density field,
$M$ is halo mass, $\bar{\rho}_{\rm m}(z)$ is the background matter
density at redshift $z$ and $\ud n(M, z)/\ud M$ is the differential
halo mass function, for the 40-512, 40-2048 and 160-2048 runs (top
panel).\footnote{Strictly speaking, the quantity shown is
  not the halo multiplicity function $f(\sigma,z)$, but is a
  closely related function, which we plot to reduce the large dynamic
  range in $\ud n /\ud M$.} In the same figure, we also show the ratio
of $\ud n/\ud M$ for 40-2048, 40-1024 and 160-2048 to 40-512 (bottom
panel), demonstrating -- for
$10^{9} \leq M_{\rm h}/\msolh \leq 10^{11}$ -- convergence to better than 20
per cent (and mostly better than 10 per cent) for all $z$ with respect
to both box size and mass resolution. The smaller volume of the
$40~\mpch$ runs means that haloes with $M_{\rm h}> 10^{11}~\msolh$ are
rarer than in the $160~\mpch$ run and, above this mass, the mass
functions are clearly not converged with box size or resolution. When
making comparisons between runs, we therefore restrict ourselves to
haloes with masses $10^{10} \leq M_{\rm h}/\msolh \leq 10^{11}$.

\begin{figure}
  \centering
  \includegraphics[width=\columnwidth]{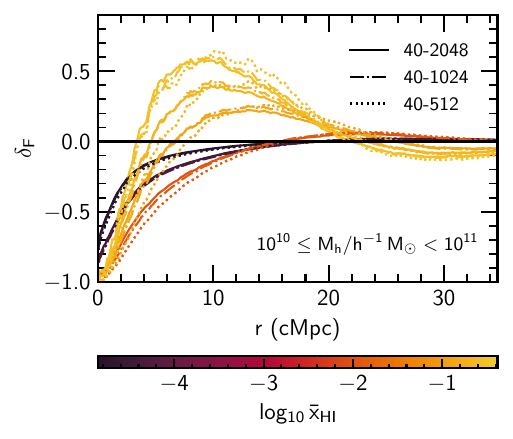}
  \vspace{-0.7cm}
  \caption{Galaxy--\lya{} transmission correlation as a function of
    distance $r$ from haloes with mass
    $10^{10} \leq M_{\rm h}/\msolh < 10^{11}$ for different
    Sherwood-Relics runs, compared at approximately equal
    volume-averaged neutral fraction $\bar{x}_{\rm HI}$. We show the
    fiducial 40-2048 (solid) simulation with the 40-1024 (dot-dashed)
    and 40-512 (dotted) simulations, which have the same box size of
    40~$\mpch$ but a particle mass that is 8 and 64 times larger,
    respectively, demonstrating the impact of mass resolution.}
  \label{fig:app-F-r}
\end{figure}

In \fref{fig:app-F-r} we show $\delta_F$ for the $40~\mpch$ Sherwood-Relics
realisation at three mass resolutions, where the 40-1024 (40-512) run
has a particle mass 8 (64) times larger than the 40-2048 run. We
compare the three runs at approximately equal volume-averaged neutral
fraction $\bar{x}_{\rm HI}$ to eliminate any differences due to
slightly different reionisation histories. We also note that each run
is normalised to the global \lya{} transmission \emph{for that run},
which is in general different between runs. We find that the form of
$\delta_F$ for our fiducial 40-2048 model is broadly converged
with mass resolution.

\section{Galaxy--transmission spike two-point cross-correlation}
\label{sec:app-2pccf}
At high redshifts ($z \gtrsim 6$), the transmission in the \lya{} forest
decreases to such a degree (see \fref{fig:xHI-all}) that it approaches
the noise level of spectrographs ($F\sim 0.02$, see e.g., fig.~8 in
\citealt{meyer2020}), making measuring $\delta_F$ an increasingly difficult
task. \citet{meyer2020} therefore proposed using a slightly different
statistic, instead measuring the two-point cross-correlation function
(2PCCF) of galaxies and \lya{} transmission spikes.  However, what
constitutes a transmission spike will depend on many factors, such as
the redshift at which the transmission is measured, or the resolution
and noise level of the instrument being used. These complications mean
we have opted to use $\delta_F$ in our main analysis, as it can be applied
more readily to disparate data sets. \citet{meyer2020} define a spike
as a pixel at $z\sim 6$ with a signal-to-noise ratio $>3$ (after
convolution with a Gaussian, see their sec.~3 for details) and
$F>0.02$. With the spikes identified, \citet{meyer2020} use the
following estimator for the 2PCCF, originally due to \citet{davis1983}
\begin{equation}
  \label{eq:2pccf}
  \xi(r) = \frac{n_{\rm R}}{n_{\rm D}}\frac{D_{\rm G}D_{\rm S}(r)}{R_{\rm G}D_{\rm S}(r)} - 1,
\end{equation}
where $D_{\rm G}D_{\rm S}$ is the number of galaxy-spike pairs as a
function of $r$, $R_{\rm G}D_{\rm S}$ is the number of random-spike
pairs (where the random positions are drawn from a uniform
distribution) as a function of $r$ and $n_{\rm G}$ and $n_{\rm R}$ are
the total numbers of galaxy-spike and random-spike pairs, used for
normalisation.

\begin{figure}
  \centering
  \includegraphics[width=\linewidth]{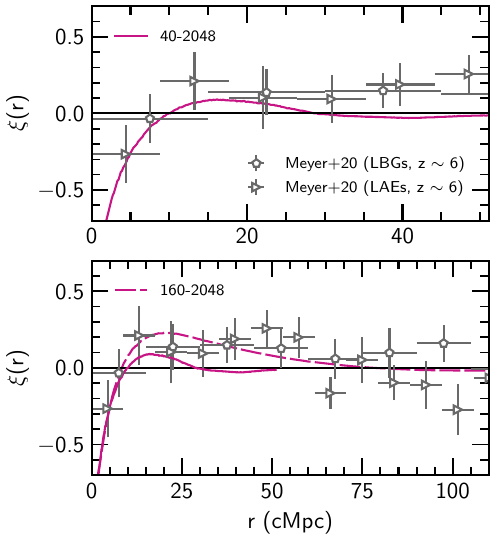}
  \vspace{-0.7cm}
  \caption{The two-point cross-correlation function between galaxies
    and transmission spikes at $z=6$ in the 40-2048 (solid, top and
    bottom panels) and 160-2048 (dashed, bottom panel)
    models. Also shown are the constraints due to \citet{meyer2020}
    (grey points). Note the different ranges of $r$ between the top
    and bottom panels.\label{fig:app-2pccf}}
\end{figure}

In \fref{fig:app-2pccf} we show the galaxy-spike 2PCCF measured from
our simulations, where we identify spikes as any pixel with $F>0.02$
and galaxies as haloes with $M_{\rm h}\geq 10^{10}~\msolh$. We do not
model observational effects (e.g., resolution or noise), deferring a
study of their impact to future work. We now find excellent agreement
between the 160-2048 model and the \citet{meyer2020} correlation with
LBGs (mostly $<1\sigma$) and slightly poorer agreement with their
correlation with LAEs. As found in \sref{sec:res-com}, for
$r \lesssim 20~{\rm cMpc}$ we find excellent agreement (always
$<1\sigma$) between the 40-2048 and both the \citet{meyer2020} LBG and LAE
correlations, but poor agreement at larger $r$.  The lack of a
redshift mismatch also suggests that the most plausible explanation
for the apparent tension with the \citet{meyer2019} $\delta_{F}$
measurements is the uncertain \civ{} absorber bias, rather than
differences in the reionisation history.


\bsp	
\label{lastpage}
\end{document}